\documentclass[aps,prl,twocolumn,floats,superscriptaddress,tighten,letterpaper,floatfix]{revtex4-1}
\usepackage[colorlinks,urlcolor=blue,anchorcolor=blue,linkcolor=blue,citecolor=blue,breaklinks=true]{hyperref}
\usepackage{amsmath}
\usepackage{amssymb}
\usepackage{graphicx}
\usepackage{color}
\usepackage{bm}
\begin{document}
\title{Topological states in a dimerized system with staggered magnetic fluxes}
\author{Ai-Lei He}
\affiliation{College of Physics Science and Technology, Yangzhou University, Yangzhou 225002, China}
\affiliation{Institute for Advanced Study, Tsinghua University, Beijing 100084, China}
\author{Wei-Wei Luo}
\affiliation{Westlake Institute of Advanced Study, Westlake University, Hangzhou, 310024, China}
\author{Yuan Zhou}
\email{zhouyuan@nju.edu.cn}
\affiliation{National Laboratory of Solid State Microstructures and Department of Physics, Nanjing University, Nanjing 210093, China}
\affiliation{Department of Mathematics and Physics, Xinjiang Teacher's College, Urumqi 830043, China}
\author{Yi-Fei Wang}
\affiliation{Center for Statistical and Theoretical Condensed Matter Physics, and Department of Physics, Zhejiang Normal University, Jinhua 321004, China}
\author{Hong Yao}
\email{yaohong@tsinghua.edu.cn}
\affiliation{Institute for Advanced Study, Tsinghua University, Beijing 100084, China}
\date{\today}
\newcommand*\mycommand[1]{\texttt{\emph{#1}}}
\newcommand{\red}[1]{\textcolor{red}{#1}}
\newcommand{\blue}[1]{\textcolor{blue}{#1}}
\newcommand{\green}[1]{\textcolor{green}{#1}}

\begin{abstract}
Robust edge states propagate along the edges and corner states gather at the corners in two-dimensional (2D) first-order and second-order topological insulators, respectively. Here, we report two kinds of topological states in an extended 2D dimerized lattice with staggered flux threading. At $\frac{1}{2}$-filling, we observe isolated corner states as well as metallic near-edge states in the $\mathcal{C}=2$ Chern insulator states. At $\frac{1}{4}$-filling, we find a $\mathcal{C}=0$ topological state, where the robust edge states are well localized along the edges but bypass corners. These topological insulator states differ from both conventional Chern insulators and the usual high-order topological insulators.
\end{abstract}
\maketitle

\section{I. Introduction}The bulk-boundary correspondence in topological states reveals the intrinsic relationship between topological invariants and boundary states~\cite{TI1,TI2}. For instance, soliton excitation with an $e/2$ fractional charge in the Su-Schrieffer-Hegger (SSH) model is the prototypical bulk-boundary correspondence in one dimension~\cite{SSH1,SSH2}. In two-dimensional (2D) topological states of matter, topologically protected boundary states propagating along the edges are robust against defects and cannot diffuse into the bulk or only gather at the corners. As a representative, Chern insulators (CIs) first proposed in the honeycomb lattice by staggered flux threading\cite{Haldane} and later experimentally realized in ultracold fermion system~\cite{Reali_HM} and magnetic topological insulators~\cite{XueQiKun} host gapped bulk bands and gapless chiral edge modes. Such bulk-edge correspondence is universal in all CIs established on various crystal lattice models, such as the checkerboard-lattice model~\cite{CB0,CB1}, the kagome-lattice model~\cite{KG1,KG2}, and other exotic lattice models~\cite{LDM,Lieb0,Star0,Star1,SQOC0,Ruby,TR}. Even in aperiodic systems, for example the recently proposed quasicrystalline CI on the D{\"u}rer's tiling~\cite{QCCI}, the bulk-edge correspondence remains applicable.

In contrast, in higher-order topological insulators (HOTIs), conventional bulk-edge correspondence no longer applies~\cite{HOTI1,HOTI2,HOTI3,HOTI4,HOTI5,HOTI6,HOTI8,HOTI9,HOTI10,HOTI11,HOTI12,HOTI13}.
HOTIs belong to a special class of topological states, hosting $(d-f)$-dimensional boundary states where $d$ labels the dimensions of the bulk and $f$ the order of HOTIs. Some 2D HOTIs are characterized by the nested Wilson loop~\cite{HOTI1,HOTI2}, reflecting the topological property of Wannier bands. As a consequence, gapped edge states and topologically protected corner states are observed, together with fractional charges at corners. There are also some unconventionally topological states hosting special bulk-boundary correspondence. Zero Chern number (i.e., $\cal C$=0), but with topologically protected edge states had been reported in the half-filled dimerized Hofstadter model~\cite{SSH4} and the quarter-filled 2D SSH model~\cite{Liu1}. So far, the boundary states observed in 2D topological insulators are either edge-like or corner-like. Whether both isolated corner and gapless edge states coexist in the same topological boundary state remains unclear.

\begin{figure}[b]
\includegraphics[scale=0.45]{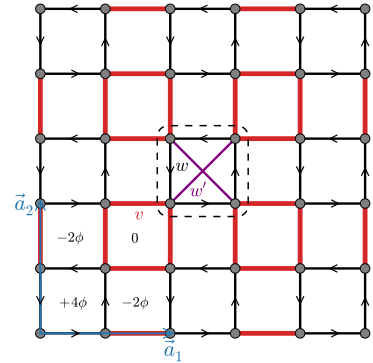}
\caption{Schematic structure of an extended SSH model on a square lattice. The lattice vectors are $\vec{a}_1$, $\vec{a}_2$ and within the unit cell, there are four atoms. Black and purple bonds represent the intracellular (the dashed square) NN ($w$) and NNN ($w^{\prime}$) hoppings, respectively, and red bonds are the intercellular hoppings ($v$). The staggered fluxes are $4\phi$ in the intercellular square, $0$ for the fully dimerized square and $-2\phi$ in the both remaining squares, respectively. The adopted gauge is explicitly displayed by arrows, with an additional phase factor $\pm \phi$ for part of the NN hopping process.}
\label{Model}
\end{figure}

In this paper, we report two kinds of topologically insulating states in a 2D dimerized square lattice model with staggered flux threading. At $1/2-$filling, we find a $\mathcal{C}=2$ Chern insulator state in which the boundary states both gather around the corners and extend into the near-edge, leading to isolated corner states and a metallic near-edge state. This topological state is also identified by Wannier bands and a significant shift of Wannier bands is observed. We argue that this behavior originates from the existence of 1D SSH-like domain walls at the corners, which isolate the corner states and push the edge states into the near-edge. On the other hand, a zero-Chern number topological insulator state emerges at $1/4$-filling. The boundary states bypass the corners and localize along the edges robustly. Our findings offer opportunities to search for more topological states.

\section{II. Model}Our starting model is an extended version of the 2D SSH model\cite{SSH1,SSH2} with staggered fluxes threading the plaquettes as illustrated in Fig.~\ref{Model}. The Hamiltonian is given by
\begin{eqnarray}
H= &&w\sum_{\langle\mathbf{r}\mathbf{r}^{\prime}\rangle}
\left[c^{\dagger}_{\mathbf{r}^{\prime}}c_{\mathbf{r}}  \exp\left(i\phi_{\mathbf{r}^{ \prime}\mathbf{r}}\right)  +\textrm{H.c.}\right]+ \nonumber\\
&&w^{\prime}\sum_{\langle\langle\mathbf{r}\mathbf{r}^{ \prime}\rangle\rangle}
\left[c^{\dagger}_{\mathbf{r}^{\prime}}c_{\mathbf{r}}  +\textrm{H.c.}\right] +
v\sum_{\langle\mathbf{r}\mathbf{r}^{\prime}\rangle^{\prime}}
\left[c^{\dagger}_{\mathbf{r}^{\prime}}c_{\mathbf{r}}+\textrm{H.c.}\right],
\label{e.1}
\end{eqnarray}
where $c^{\dagger}_{\mathbf{r}}$ ($c_{\mathbf{r}}$) is the creation (annihilation) operator of a spinless electron at site $\mathbf{r}$. $\langle\cdots\rangle$ and $\langle\langle\cdots\rangle\rangle$ denote the intracellular nearest-neighbor (NN), and the next-nearest-neighbor (NNN) hopping process with hopping integrals $w$, and $w^{\prime}$, respectively. $\langle\cdots\rangle^{\prime}$ denotes the intercellular NN hopping with the amplitude $v$. $\phi_{\mathbf{r}^{ \prime}\mathbf{r}}=\pm\phi$ is the phase factor in the intracellular NN hopping process with the sign denoted by the arrow. Numerically, we set the intracellular NN hopping integral $w$ as a unit, $w^{\prime}=0.75$, and $\phi=-0.25\pi$ unless specified. The present model reduces to the $2$D SSH model in the absence of the flux and NNN hopping process, where a Zak phase accompanied by the fractional wave polarization with vanishing Berry curvature in the first Brillouin zone is revealed~\cite{Liu1}. A similar dimerized $2$D lattice model with $\pi$ flux threading hosts a HOTI in which the robust corner states with a fractional corner charge $e/2$ emerge due to the quantized quadrupole moments~\cite{HOTI1,HOTI2}.

\begin{figure}[!htb]
\includegraphics[scale=0.73]{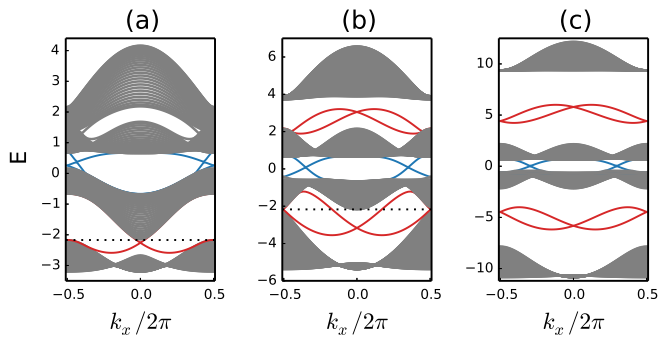}
\caption{Topological band structures and edge states in the model Hamiltonian on the cylinder geometry with, (a) $v=1.0$, (b) $v=2.2$, and (c) $v=5.0$. The edge states are highlighted by colors. }
\label{EB}
\end{figure}

We show the energy bands evolution on the intercellular hopping $\nu$ on the cylinder geometry (shown in Fig.~\ref{EB}). There are four bands in our model. A significant band gap between the upper two and lower two bands emerges with two chiral edge modes, suggesting a CI with $\mathcal{C}=2$ at half-filling. On the other hand, the band gap between the two lowest bands evolves from an indirect gap to a direct gap with increasing $v$, indicating a metallic to insulating phase transition.

\begin{figure*}
\centering
\includegraphics[scale=0.85]{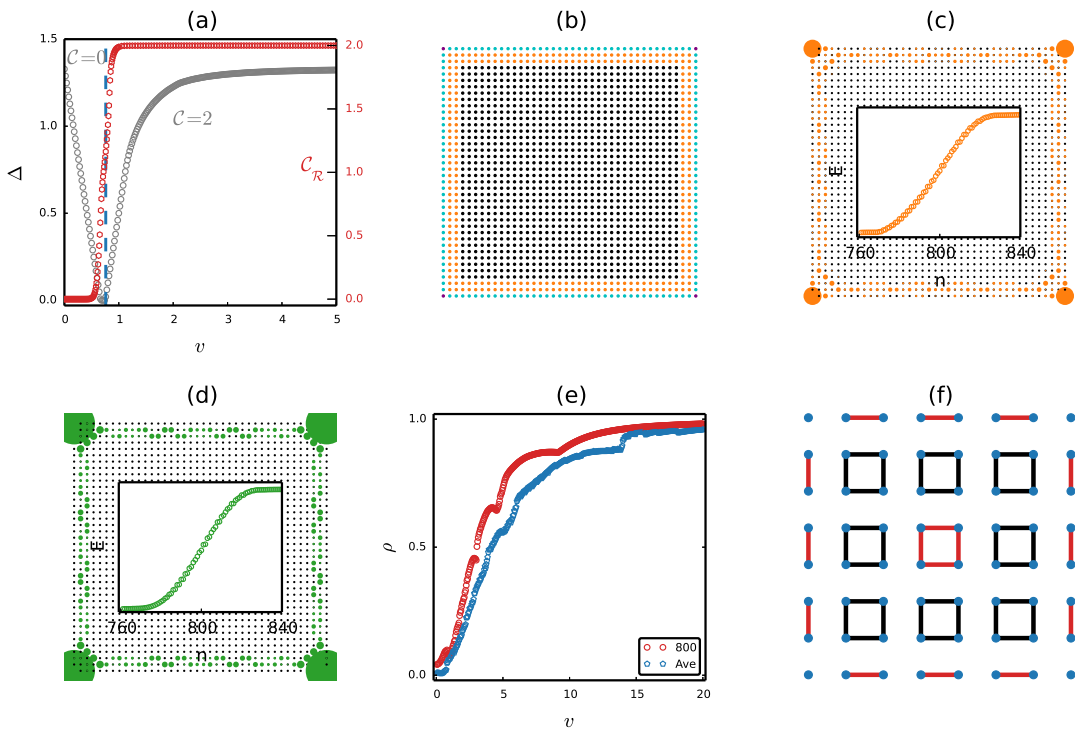}
\caption{  (a) Energy band gap $\Delta$ on the torus geometry and real-space Chern number ${\cal C}_R$ vs $v$ in a plane with defects.  (b) We divide a 1600-sites square lattice into four parts, the corners (purple), the edge except for corners (cyan), the near-edge (orange) and the bulk (black), respectively. Spatial wavefunction of the present CI state (the $800$th state) shown in (c), and (d) with $v= 2.2$ and $5.0$. The insets display energy eigenvalues $E$ vs the state index $n$. (e). The ratio of corner density ($\rho$) of the $800$th in-gap state, together with the average of 40 in-gap states around the $800$th in-gap state (Ave). (f). Illustration of the model at strong dimerization limitation, i.e. $v=0$. We mark the strong bonds with different colors. }
\label{en_wf}
\end{figure*}

\section{III. ${\cal C}=2$ Chern insulator states at $1/2$-filling}The topological property at $1/2$-filling can be identified by the Chern number $\mathcal{C}=\sum_{E_{n}<E_{F}}\mathcal{C}_{n}$ on the torus geometry. Here, $E_{F}$ is the Fermi energy and $\mathcal{C}_{n}$ is the sub-band Chern number calculated by integrating the Berry curvature over the first Brillouin zone~\cite{TKNN,KG2,suppl}. Numerical results show that the model hosts phase transitions from a trivial state with $\mathcal{C}=0$ to a topologically nontrivial state with $\mathcal{C}=2$ at the critical point $v=0.75$, which can also be identified with the aid of the band gap $\Delta$ [Fig.~\ref{en_wf}(a)]. This Chern insulator state agrees with the spectrum on the cylinder geometry, where two branches of chiral edge states appear in the band gap.

To account for the topological properties of the boundary states, we divide the lattice with an open boundary into four parts---the corners, the edges (excluding corners), the near-edge, and the bulk~[as illustrated in Fig.~\ref{en_wf} (b)]. Spatial distributions of the boundary states are shown in Figs.~\ref{en_wf} (c) and ~\ref{en_wf} (d). Interestingly, the corner states are gradually enhanced while the edge states are weakened with increasing $v$~\cite{suppl}. At a stronger dimerized potential ($v>2.2$), the edge states slightly diffuse into the bulk, leaving the edges empty and the corner states isolated---we call them the near-edge states. These features are in sharp contrast to the well edge-localized boundary states in the conventional CIs~\cite{suppl}, suggesting the special topological features of the boundary states at half-filling. We define the ratio of the corner density as $\rho = \vert\phi^n_{\text{cor}}\vert^2/ \vert\phi^n_{\text{bd}}\vert^2 $ to measure the proportion of the corner states in the boundary states (including corners and edges), where $\vert\phi^n_{\text{cor}}\vert^2$ ($\vert\phi^n_{\text{bd}}\vert^2$) denotes the corner (boundary) density of the $n$-th state. The ratio $\rho$ of the $800$th state enhances with the increasing $\nu$ [Fig.~\ref{en_wf} (e)]. At a strong dimerizaton limitation, the ratio is nearly equal to $1.0$, which suggests the boundary states are fully gapped. In fact, the in-gap states host similar tendency as shown in the average ratio of the corner density from the $780$th to the $820$th in-gap states.

The coexistence of corner and near-edge states is robust against the strong intercellular hopping in the present topological state, for example, $v=20$ and $v=50$~\cite{suppl}. These special boundary states are also robust against defects and lattice sizes~\cite{suppl}. In order to present the topological features in an open system with defects, we further compute the real-space Chern number based on the Kitaev formula~\cite{LCHN1,suppl}. In Fig.~\ref{en_wf} (a) (red line), we present the real-space Chern number with a selected Fermi energy at the $795$th energy level in a 1588-sites lattice with defects [for details see the Supplemental Material (SM)]. The real-space Chern number is near two when $v>0.80$, which is slightly larger than the critical value of $0.75$ on the torus geometry due to the finite-size effect.  We emphasize that no fractional charge at the corners emerges even at strong dimerization limitation (for details see SM~\cite{suppl}), suggesting the CI rather than the HOTI nature~\cite{HOTI1,HOTI2} of the present topological state.

\begin{figure}[t]
\includegraphics[scale=0.7]{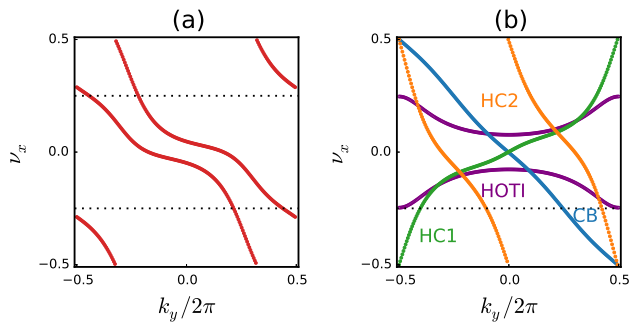}
\caption{(a) Wannier spectra of the present CI state. (b) As a reference, we calculate the Wannier centers of a Haldane model with Chern number -1 (CH1), Chern number 2 (CH2)~\cite{HC2}, checkerboard model (CB), and HOTI~\cite{HOTI1}. Here dashed lines are the reference line.}
\label{WCs}
\end{figure}

In some $2$D HOTIs, topological properties are well characterized by a topological invariant, a nested Wilson loop~\cite{HOTI1,HOTI2}, where  gapped Wannier bands are discovered. On the contrary, Wannier bands are gapless in conventional CIs~\cite{HOTI2,HOTI13}, consisting  of  gapless edge states. Here, we further study Wannier bands to identify the present CI state. The Wannier spectra are shown in Fig.~\ref{WCs} (a). The gapless Wannier bands in our model manifest the CI nature of the present topological state. Wannier centers crossing a reference line twice further confirm the $\mathcal{C}=2$ Chern insulator state. Such features are in sharp contrast to the gapped Wannier bands in a HOTI [Fig.~\ref{WCs} (b)]~\cite{HOTI1,HOTI2}. On the other hand, unlike the exact value $\nu_{x}=\pm 1/2$ at $k_{y}=\pm \pi$ in conventional CI states [such as the Haldane model\cite{Haldane,HC2} and the checkerboard model\cite{CB1} in Fig.~\ref{WCs} (b)], significant Wannier band shifting is observed in our present model. This agrees well with the special features of the isolated corner states and the near-edge states.

We explain the coexistence of the corner and near-edge states by considering the dimerized bond limitation, $i.e., v=0$. In this limiting case, we find four isolated atoms at the corners where some states are localized [details shown in Fig.~\ref{en_wf} (f)] and no state propagates along the edge, $i.e.$ the gapped boundary state. Concomitantly, the gapped boundary pushes the gapless edge states inherited from the CI nature into the near-edges and the topologically protected state propagates along the near-edge bonds marked with black in Fig.~\ref{en_wf} (f). On the other hand, we can explain  the coexistence of corner and near-edge states based on the 1D SSH-like domain walls at the corners. At a weak dimerization, the domain wall remains trivial and the system exhibits a conventional CI nature with robust edge states~\cite{suppl}. On the contrary, at strong dimerization, nontrivial domain walls yield gapped boundary states and isolated corner states but without fractional charge. This gapped boundary pushes the gapless edge states into near-edges because of the CI nature.

\section{IV. ${\cal C}=0$ topological state at $1/4$-filling}
\begin{figure}[t]
\includegraphics[scale=0.68]{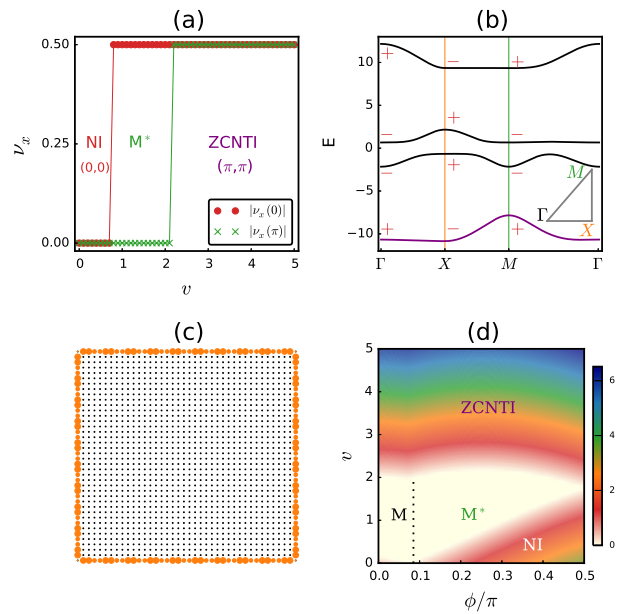}
\caption{(a). At $1/4$-filling, three states, including the normal insulator (NI) [with 2D Zak phase (0,0)], the metallic (M$^*$) and the zero-Chern number topological insulator (ZCNTI) phases [with 2D Zak phase ($\pi$,$\pi$)], are observed. $\nu_x$ denotes the value of the Wannier bands. (b). Bulk bands of our present model with parities at high  symmetric points ($\Gamma$, $X$ and $M$) as ``$\pm$". (c) Single-particle state of the zero-Chern number topological state. (d). Phase diagram of the model at $1/4$-filling obtained based on the energy gap. There are three phase regions, the zero-Chern number topological insulator, the metal and the normal insulator. The metal phase contain two types of metals. One is the normal metal (M) and the other is with CI-like features (M*). In (b) and (c), we set $v=5.0$.}
\label{fra_c}
\end{figure}

We now study the topological properties of the proposed model at $1/4$-filling. By tuning the intercellular hopping potential, chiral and helical-like edge states emerge (in Fig.~\ref{EB}), which suggests the appearance of various topological states. We analyze their Wannier centers on the torus geometry to identify these topological states [in Fig.~\ref{fra_c} (a)]. In the case of weak intercellular hopping $v<0.75$, the Wannier bands show a typically normal insulating feature, i.e., $\nu_{x}(0)=\nu_{x}(\pi)=\nu_{x}(2\pi)\equiv0$, as well the near zero value for the other momentum. When $v\in(0.75, 2.2)$, a CI-like feature of the gapless Wannier band with $\nu_{x}(0)=0.5$, $\nu_{x}(\pi)=0$, and $\nu_{x}(2\pi)=-0.5$ emerges in the intermediate $v$~\cite{suppl}. However, due to the indirect band gap between the lowest two bands shown in Fig.~\ref{EB} (a), the present state is a metallic state.

Further increasing $v>2.2$, the reopening band gap (in Fig.~\ref{EB}), together with specific Wannier centers indicate the emergence of another topological state [Fig.~\ref{fra_c} (a)]. It should be reminded that there are two isolated edge modes in the band gap propagating reversely along the open boundaries [Fig.~\ref{EB} (c)]. On account of our model with inversion symmetry, the topology of the energy band can be further confirmed by the $2$D Zak phase based on the parities at high symmetric points~\cite{ISTIs,CFang,SSH4,Liu1,suppl}, which are marked as ``$\pm$" displayed in Fig.~\ref{fra_c} (b). We obtain the polarization ${\bf p}=(1/2, 1/2)$, which is in accordance with the results of the Wannier centers [shown in Fig.~\ref{fra_c} (a)]. We find this topological state and the previously reported zero Berry curvature topological state in the $2$D SSH model~\cite{Liu1} belong to the same topological class based on the bulk energies~\cite{suppl}. The robustness of this topological state is solely determined by the hopping parameters. One feature of this topological state is that the edge states disconnect the valence and conduction bands because this edge state strongly depends on the $C_{4v}$ point group symmetry and the open boundary condition breaks the $C_{4v}$ symmetry~\cite{Liu1}. Reference~\cite{Liu1} has pointed out that the edge states connect the valence and conduction bands when a nontrivial ribbon is bounded by another trivial ribbon with $C_{4v}$ symmetry. In addition, we find this topological state is not protected by inversion symmetry because the edge states still appear when the inversion symmetry is broken~\cite{suppl}.

The spatial distribution of the boundary state is presented in Fig.~\ref{fra_c} (c). The boundary state is well localized along the edges but bypasses the corners, leaving the missing of corner states. This topological state is contrary to the conventional topological states, in which both the edge and corner states exist. On the other hand, it also differs from the $2$D HOTI states, where only corner states emerge. At 1/2-filling, a partial boundary state is localized around the isolated corners and no fractional charge accumulates around every corner. Therefore, at 1/4-filling, the boundary state bypasses the corners based on the charge conservation near the corners. Simultaneously, an $e/4$ fractional charge emerges around every corners~\cite{suppl} and it has been reported in the 2-D SSH model~\cite{polarcharge4}.  We also present the phase diagram by tuning the intercellular hopping $v$ and  staggered flux $\phi$ [Fig.~\ref{fra_c} (d)], which contains the normal (trivial) insulator, the metal and the zero-Chern number topological phases. We notice there are no CI phase, because the present parameters are not able to absolutely open the band gap at $1/4-$filling and finally lead to a metal phase with CI-like features [Fig.~\ref{fra_c} (d)]. In addition, this zero-Chern number topological state can emerge at 3/4-filling as well. For example, when $v=5.0$, we can obtain the edge state [highlighted in Fig.~\ref{EB} (c)] and the quantized polarization ${\bf p}=(1/2, 1/2)$ which is calculated  based on the parities of the band at 3/4-filling [details are shown in Fig.~\ref{fra_c}].

\section{V. Summary and discussion}We have proposed two kinds of topological insulator states realized in a dimerized square lattice model with staggered fluxes threading. At half-filling, the $\mathcal{C}=2$ CI state, identified by the Chern number and the gapless Wannier bands,  coexists with isolated corner states and the near-edge states.  We argue these special boundary states come from the existence of a $1$D SSH-like domain wall. At quarter-filling, we find another topological insulator state with $\mathcal{C}=0$. The boundary state is localized along the edges but in the absence of the corner state. Our findings may add insight into different boundary states in topological insulators.

Experimentally, a very promising approach to realize dimerized quantum Hall states is the ultracold atomic system, in which the Hofstadter model~\cite{Hofs} and the Haldane model~\cite{Reali_HM} had been realized. Recently, $2$D dimerized lattices hosting HOTIs have been theoretically discussed and experimentally observed in phononic and electrical-circuit systems~\cite{HOTI1,Phot_HOTI,Phot_HI,HOTI11,Circ_HOTI,Peterson}, and CI states have been realized in electrical-circuit systems~\cite{Circ_chern}. Hence, based on those potential ways to construct dimerized lattices and simulate CI states, it would be natural to make experimental observation of more topological states in various artificial microstructures. The observation of a CI state in a multilayer graphene moir{\'e} superlattice~\cite{TGM,TBG_MH,SC_TBG} also creates a possible route to explore our present topological states.

{\it Acknowledgments.---}
We thank to Y.Y. Huang for helpful discussions. This work was supported in part by the NSFC under Grants No. 11825404 (A.-L.H. and H.Y.), No. 11874325 (Y.-F.W.) and No. 12074175 (Y. Z.), the MOSTC under Grants No. 2016YFA0300401 (Y.Z.) and No. 2018YFA0305604 (H.Y.), and the Strategic Priority Research Program of Chinese Academy of Sciences under Grant No. XDB28000000 (H.Y.).

\bibliography{ACI}

\begin{thebibliography}{49}%
\makeatletter
\providecommand \@ifxundefined [1]{%
 \@ifx{#1\undefined}
}%
\providecommand \@ifnum [1]{%
 \ifnum #1\expandafter \@firstoftwo
 \else \expandafter \@secondoftwo
 \fi
}%
\providecommand \@ifx [1]{%
 \ifx #1\expandafter \@firstoftwo
 \else \expandafter \@secondoftwo
 \fi
}%
\providecommand \natexlab [1]{#1}%
\providecommand \enquote  [1]{``#1''}%
\providecommand \bibnamefont  [1]{#1}%
\providecommand \bibfnamefont [1]{#1}%
\providecommand \citenamefont [1]{#1}%
\providecommand \href@noop [0]{\@secondoftwo}%
\providecommand \href [0]{\begingroup \@sanitize@url \@href}%
\providecommand \@href[1]{\@@startlink{#1}\@@href}%
\providecommand \@@href[1]{\endgroup#1\@@endlink}%
\providecommand \@sanitize@url [0]{\catcode `\\12\catcode `\$12\catcode
  `\&12\catcode `\#12\catcode `\^12\catcode `\_12\catcode `\%12\relax}%
\providecommand \@@startlink[1]{}%
\providecommand \@@endlink[0]{}%
\providecommand \url  [0]{\begingroup\@sanitize@url \@url }%
\providecommand \@url [1]{\endgroup\@href {#1}{\urlprefix }}%
\providecommand \urlprefix  [0]{URL }%
\providecommand \Eprint [0]{\href }%
\providecommand \doibase [0]{http://dx.doi.org/}%
\providecommand \selectlanguage [0]{\@gobble}%
\providecommand \bibinfo  [0]{\@secondoftwo}%
\providecommand \bibfield  [0]{\@secondoftwo}%
\providecommand \translation [1]{[#1]}%
\providecommand \BibitemOpen [0]{}%
\providecommand \bibitemStop [0]{}%
\providecommand \bibitemNoStop [0]{.\EOS\space}%
\providecommand \EOS [0]{\spacefactor3000\relax}%
\providecommand \BibitemShut  [1]{\csname bibitem#1\endcsname}%
\let\auto@bib@innerbib\@empty
\bibitem [{\citenamefont {Hasan}\ and\ \citenamefont {Kane}(2010)}]{TI1}%
  \BibitemOpen
  \bibfield  {author} {\bibinfo {author} {\bibfnamefont {M.~Z.}\ \bibnamefont
  {Hasan}}\ and\ \bibinfo {author} {\bibfnamefont {C.~L.}\ \bibnamefont
  {Kane}},\ }\href {\doibase 10.1103/RevModPhys.82.3045} {\bibfield  {journal}
  {\bibinfo  {journal} {Rev. Mod. Phys.}\ }\textbf {\bibinfo {volume} {82}},\
  \bibinfo {pages} {3045} (\bibinfo {year} {2010})}\BibitemShut {NoStop}%
\bibitem [{\citenamefont {Qi}\ and\ \citenamefont {Zhang}(2011)}]{TI2}%
  \BibitemOpen
  \bibfield  {author} {\bibinfo {author} {\bibfnamefont {X.-L.}\ \bibnamefont
  {Qi}}\ and\ \bibinfo {author} {\bibfnamefont {S.-C.}\ \bibnamefont {Zhang}},\
  }\href {\doibase 10.1103/RevModPhys.83.1057} {\bibfield  {journal} {\bibinfo
  {journal} {Rev. Mod. Phys.}\ }\textbf {\bibinfo {volume} {83}},\ \bibinfo
  {pages} {1057} (\bibinfo {year} {2011})}\BibitemShut {NoStop}%
\bibitem [{\citenamefont {Su}\ \emph {et~al.}(1979)\citenamefont {Su},
  \citenamefont {Schrieffer},\ and\ \citenamefont {Heeger}}]{SSH1}%
  \BibitemOpen
  \bibfield  {author} {\bibinfo {author} {\bibfnamefont {W.~P.}\ \bibnamefont
  {Su}}, \bibinfo {author} {\bibfnamefont {J.~R.}\ \bibnamefont {Schrieffer}},
  \ and\ \bibinfo {author} {\bibfnamefont {A.~J.}\ \bibnamefont {Heeger}},\
  }\href {\doibase 10.1103/PhysRevLett.42.1698} {\bibfield  {journal} {\bibinfo
   {journal} {Phys. Rev. Lett.}\ }\textbf {\bibinfo {volume} {42}},\ \bibinfo
  {pages} {1698} (\bibinfo {year} {1979})}\BibitemShut {NoStop}%
\bibitem [{\citenamefont {Su}\ \emph {et~al.}(1980)\citenamefont {Su},
  \citenamefont {Schrieffer},\ and\ \citenamefont {Heeger}}]{SSH2}%
  \BibitemOpen
  \bibfield  {author} {\bibinfo {author} {\bibfnamefont {W.~P.}\ \bibnamefont
  {Su}}, \bibinfo {author} {\bibfnamefont {J.~R.}\ \bibnamefont {Schrieffer}},
  \ and\ \bibinfo {author} {\bibfnamefont {A.~J.}\ \bibnamefont {Heeger}},\
  }\href {\doibase 10.1103/PhysRevB.22.2099} {\bibfield  {journal} {\bibinfo
  {journal} {Phys. Rev. B}\ }\textbf {\bibinfo {volume} {22}},\ \bibinfo
  {pages} {2099} (\bibinfo {year} {1980})}\BibitemShut {NoStop}%
\bibitem [{\citenamefont {Haldane}(1988)}]{Haldane}%
  \BibitemOpen
  \bibfield  {author} {\bibinfo {author} {\bibfnamefont {F.~D.~M.}\
  \bibnamefont {Haldane}},\ }\href {\doibase 10.1103/PhysRevLett.61.2015}
  {\bibfield  {journal} {\bibinfo  {journal} {Phys. Rev. Lett.}\ }\textbf
  {\bibinfo {volume} {61}},\ \bibinfo {pages} {2015} (\bibinfo {year}
  {1988})}\BibitemShut {NoStop}%
\bibitem [{\citenamefont {Jotzu}\ \emph {et~al.}(2014)\citenamefont {Jotzu},
  \citenamefont {Messer}, \citenamefont {Desbuquois}, \citenamefont {Lebrat},
  \citenamefont {Uehlinger}, \citenamefont {Greif},\ and\ \citenamefont
  {Esslinger}}]{Reali_HM}%
  \BibitemOpen
  \bibfield  {author} {\bibinfo {author} {\bibfnamefont {G.}~\bibnamefont
  {Jotzu}}, \bibinfo {author} {\bibfnamefont {M.}~\bibnamefont {Messer}},
  \bibinfo {author} {\bibfnamefont {R.}~\bibnamefont {Desbuquois}}, \bibinfo
  {author} {\bibfnamefont {M.}~\bibnamefont {Lebrat}}, \bibinfo {author}
  {\bibfnamefont {T.}~\bibnamefont {Uehlinger}}, \bibinfo {author}
  {\bibfnamefont {D.}~\bibnamefont {Greif}}, \ and\ \bibinfo {author}
  {\bibfnamefont {T.}~\bibnamefont {Esslinger}},\ }\href
  {https://doi.org/10.1038/nature13915} {\bibfield  {journal} {\bibinfo
  {journal} {Nature}\ }\textbf {\bibinfo {volume} {515}},\ \bibinfo {pages}
  {237 EP } (\bibinfo {year} {2014})}\BibitemShut {NoStop}%
\bibitem [{\citenamefont {Chang}\ \emph {et~al.}(2013)\citenamefont {Chang},
  \citenamefont {Zhang}, \citenamefont {Feng}, \citenamefont {Shen},
  \citenamefont {Zhang}, \citenamefont {Guo}, \citenamefont {Li}, \citenamefont
  {Ou}, \citenamefont {Wei}, \citenamefont {Wang}, \citenamefont {Ji},
  \citenamefont {Feng}, \citenamefont {Ji}, \citenamefont {Chen}, \citenamefont
  {Jia}, \citenamefont {Dai}, \citenamefont {Fang}, \citenamefont {Zhang},
  \citenamefont {He}, \citenamefont {Wang}, \citenamefont {Lu}, \citenamefont
  {Ma},\ and\ \citenamefont {Xue}}]{XueQiKun}%
  \BibitemOpen
  \bibfield  {author} {\bibinfo {author} {\bibfnamefont {C.-Z.}\ \bibnamefont
  {Chang}}, \bibinfo {author} {\bibfnamefont {J.}~\bibnamefont {Zhang}},
  \bibinfo {author} {\bibfnamefont {X.}~\bibnamefont {Feng}}, \bibinfo {author}
  {\bibfnamefont {J.}~\bibnamefont {Shen}}, \bibinfo {author} {\bibfnamefont
  {Z.}~\bibnamefont {Zhang}}, \bibinfo {author} {\bibfnamefont
  {M.}~\bibnamefont {Guo}}, \bibinfo {author} {\bibfnamefont {K.}~\bibnamefont
  {Li}}, \bibinfo {author} {\bibfnamefont {Y.}~\bibnamefont {Ou}}, \bibinfo
  {author} {\bibfnamefont {P.}~\bibnamefont {Wei}}, \bibinfo {author}
  {\bibfnamefont {L.-L.}\ \bibnamefont {Wang}}, \bibinfo {author}
  {\bibfnamefont {Z.-Q.}\ \bibnamefont {Ji}}, \bibinfo {author} {\bibfnamefont
  {Y.}~\bibnamefont {Feng}}, \bibinfo {author} {\bibfnamefont {S.}~\bibnamefont
  {Ji}}, \bibinfo {author} {\bibfnamefont {X.}~\bibnamefont {Chen}}, \bibinfo
  {author} {\bibfnamefont {J.}~\bibnamefont {Jia}}, \bibinfo {author}
  {\bibfnamefont {X.}~\bibnamefont {Dai}}, \bibinfo {author} {\bibfnamefont
  {Z.}~\bibnamefont {Fang}}, \bibinfo {author} {\bibfnamefont {S.-C.}\
  \bibnamefont {Zhang}}, \bibinfo {author} {\bibfnamefont {K.}~\bibnamefont
  {He}}, \bibinfo {author} {\bibfnamefont {Y.}~\bibnamefont {Wang}}, \bibinfo
  {author} {\bibfnamefont {L.}~\bibnamefont {Lu}}, \bibinfo {author}
  {\bibfnamefont {X.-C.}\ \bibnamefont {Ma}}, \ and\ \bibinfo {author}
  {\bibfnamefont {Q.-K.}\ \bibnamefont {Xue}},\ }\href {\doibase
  10.1126/science.1234414} {\bibfield  {journal} {\bibinfo  {journal}
  {Science}\ }\textbf {\bibinfo {volume} {340}},\ \bibinfo {pages} {167}
  (\bibinfo {year} {2013})}\BibitemShut {NoStop}%
\bibitem [{\citenamefont {Yakovenko}(1990)}]{CB0}%
  \BibitemOpen
  \bibfield  {author} {\bibinfo {author} {\bibfnamefont {V.~M.}\ \bibnamefont
  {Yakovenko}},\ }\href {\doibase 10.1103/PhysRevLett.65.251} {\bibfield
  {journal} {\bibinfo  {journal} {Phys. Rev. Lett.}\ }\textbf {\bibinfo
  {volume} {65}},\ \bibinfo {pages} {251} (\bibinfo {year} {1990})}\BibitemShut
  {NoStop}%
\bibitem [{\citenamefont {Sun}\ \emph {et~al.}(2011)\citenamefont {Sun},
  \citenamefont {Gu}, \citenamefont {Katsura},\ and\ \citenamefont
  {Das~Sarma}}]{CB1}%
  \BibitemOpen
  \bibfield  {author} {\bibinfo {author} {\bibfnamefont {K.}~\bibnamefont
  {Sun}}, \bibinfo {author} {\bibfnamefont {Z.}~\bibnamefont {Gu}}, \bibinfo
  {author} {\bibfnamefont {H.}~\bibnamefont {Katsura}}, \ and\ \bibinfo
  {author} {\bibfnamefont {S.}~\bibnamefont {Das~Sarma}},\ }\href {\doibase
  10.1103/PhysRevLett.106.236803} {\bibfield  {journal} {\bibinfo  {journal}
  {Phys. Rev. Lett.}\ }\textbf {\bibinfo {volume} {106}},\ \bibinfo {pages}
  {236803} (\bibinfo {year} {2011})}\BibitemShut {NoStop}%
\bibitem [{\citenamefont {Guo}\ and\ \citenamefont {Franz}(2009)}]{KG1}%
  \BibitemOpen
  \bibfield  {author} {\bibinfo {author} {\bibfnamefont {H.-M.}\ \bibnamefont
  {Guo}}\ and\ \bibinfo {author} {\bibfnamefont {M.}~\bibnamefont {Franz}},\
  }\href {\doibase 10.1103/PhysRevB.80.113102} {\bibfield  {journal} {\bibinfo
  {journal} {Phys. Rev. B}\ }\textbf {\bibinfo {volume} {80}},\ \bibinfo
  {pages} {113102} (\bibinfo {year} {2009})}\BibitemShut {NoStop}%
\bibitem [{\citenamefont {Tang}\ \emph {et~al.}(2011)\citenamefont {Tang},
  \citenamefont {Mei},\ and\ \citenamefont {Wen}}]{KG2}%
  \BibitemOpen
  \bibfield  {author} {\bibinfo {author} {\bibfnamefont {E.}~\bibnamefont
  {Tang}}, \bibinfo {author} {\bibfnamefont {J.-W.}\ \bibnamefont {Mei}}, \
  and\ \bibinfo {author} {\bibfnamefont {X.-G.}\ \bibnamefont {Wen}},\ }\href
  {\doibase 10.1103/PhysRevLett.106.236802} {\bibfield  {journal} {\bibinfo
  {journal} {Phys. Rev. Lett.}\ }\textbf {\bibinfo {volume} {106}},\ \bibinfo
  {pages} {236802} (\bibinfo {year} {2011})}\BibitemShut {NoStop}%
\bibitem [{\citenamefont {Qi}\ \emph {et~al.}(2006)\citenamefont {Qi},
  \citenamefont {Wu},\ and\ \citenamefont {Zhang}}]{LDM}%
  \BibitemOpen
  \bibfield  {author} {\bibinfo {author} {\bibfnamefont {X.-L.}\ \bibnamefont
  {Qi}}, \bibinfo {author} {\bibfnamefont {Y.-S.}\ \bibnamefont {Wu}}, \ and\
  \bibinfo {author} {\bibfnamefont {S.-C.}\ \bibnamefont {Zhang}},\ }\href
  {\doibase 10.1103/PhysRevB.74.085308} {\bibfield  {journal} {\bibinfo
  {journal} {Phys. Rev. B}\ }\textbf {\bibinfo {volume} {74}},\ \bibinfo
  {pages} {085308} (\bibinfo {year} {2006})}\BibitemShut {NoStop}%
\bibitem [{\citenamefont {Weeks}\ and\ \citenamefont {Franz}(2010)}]{Lieb0}%
  \BibitemOpen
  \bibfield  {author} {\bibinfo {author} {\bibfnamefont {C.}~\bibnamefont
  {Weeks}}\ and\ \bibinfo {author} {\bibfnamefont {M.}~\bibnamefont {Franz}},\
  }\href {\doibase 10.1103/PhysRevB.82.085310} {\bibfield  {journal} {\bibinfo
  {journal} {Phys. Rev. B}\ }\textbf {\bibinfo {volume} {82}},\ \bibinfo
  {pages} {085310} (\bibinfo {year} {2010})}\BibitemShut {NoStop}%
\bibitem [{\citenamefont {Yao}\ and\ \citenamefont {Kivelson}(2007)}]{Star0}%
  \BibitemOpen
  \bibfield  {author} {\bibinfo {author} {\bibfnamefont {H.}~\bibnamefont
  {Yao}}\ and\ \bibinfo {author} {\bibfnamefont {S.~A.}\ \bibnamefont
  {Kivelson}},\ }\href {\doibase 10.1103/PhysRevLett.99.247203} {\bibfield
  {journal} {\bibinfo  {journal} {Phys. Rev. Lett.}\ }\textbf {\bibinfo
  {volume} {99}},\ \bibinfo {pages} {247203} (\bibinfo {year}
  {2007})}\BibitemShut {NoStop}%
\bibitem [{\citenamefont {R\"uegg}\ \emph {et~al.}(2010)\citenamefont
  {R\"uegg}, \citenamefont {Wen},\ and\ \citenamefont {Fiete}}]{Star1}%
  \BibitemOpen
  \bibfield  {author} {\bibinfo {author} {\bibfnamefont {A.}~\bibnamefont
  {R\"uegg}}, \bibinfo {author} {\bibfnamefont {J.}~\bibnamefont {Wen}}, \ and\
  \bibinfo {author} {\bibfnamefont {G.~A.}\ \bibnamefont {Fiete}},\ }\href
  {\doibase 10.1103/PhysRevB.81.205115} {\bibfield  {journal} {\bibinfo
  {journal} {Phys. Rev. B}\ }\textbf {\bibinfo {volume} {81}},\ \bibinfo
  {pages} {205115} (\bibinfo {year} {2010})}\BibitemShut {NoStop}%
\bibitem [{\citenamefont {Kargarian}\ and\ \citenamefont
  {Fiete}(2010)}]{SQOC0}%
  \BibitemOpen
  \bibfield  {author} {\bibinfo {author} {\bibfnamefont {M.}~\bibnamefont
  {Kargarian}}\ and\ \bibinfo {author} {\bibfnamefont {G.~A.}\ \bibnamefont
  {Fiete}},\ }\href {\doibase 10.1103/PhysRevB.82.085106} {\bibfield  {journal}
  {\bibinfo  {journal} {Phys. Rev. B}\ }\textbf {\bibinfo {volume} {82}},\
  \bibinfo {pages} {085106} (\bibinfo {year} {2010})}\BibitemShut {NoStop}%
\bibitem [{\citenamefont {Hu}\ \emph {et~al.}(2011)\citenamefont {Hu},
  \citenamefont {Kargarian},\ and\ \citenamefont {Fiete}}]{Ruby}%
  \BibitemOpen
  \bibfield  {author} {\bibinfo {author} {\bibfnamefont {X.}~\bibnamefont
  {Hu}}, \bibinfo {author} {\bibfnamefont {M.}~\bibnamefont {Kargarian}}, \
  and\ \bibinfo {author} {\bibfnamefont {G.~A.}\ \bibnamefont {Fiete}},\ }\href
  {\doibase 10.1103/PhysRevB.84.155116} {\bibfield  {journal} {\bibinfo
  {journal} {Phys. Rev. B}\ }\textbf {\bibinfo {volume} {84}},\ \bibinfo
  {pages} {155116} (\bibinfo {year} {2011})}\BibitemShut {NoStop}%
\bibitem [{\citenamefont {Wang}\ \emph {et~al.}(2012)\citenamefont {Wang},
  \citenamefont {Yao}, \citenamefont {Gong},\ and\ \citenamefont {Sheng}}]{TR}%
  \BibitemOpen
  \bibfield  {author} {\bibinfo {author} {\bibfnamefont {Y.-F.}\ \bibnamefont
  {Wang}}, \bibinfo {author} {\bibfnamefont {H.}~\bibnamefont {Yao}}, \bibinfo
  {author} {\bibfnamefont {C.-D.}\ \bibnamefont {Gong}}, \ and\ \bibinfo
  {author} {\bibfnamefont {D.~N.}\ \bibnamefont {Sheng}},\ }\href {\doibase
  10.1103/PhysRevB.86.201101} {\bibfield  {journal} {\bibinfo  {journal} {Phys.
  Rev. B}\ }\textbf {\bibinfo {volume} {86}},\ \bibinfo {pages} {201101}
  (\bibinfo {year} {2012})}\BibitemShut {NoStop}%
\bibitem [{\citenamefont {He}\ \emph {et~al.}(2019)\citenamefont {He},
  \citenamefont {Ding}, \citenamefont {Zhou}, \citenamefont {Wang},\ and\
  \citenamefont {Gong}}]{QCCI}%
  \BibitemOpen
  \bibfield  {author} {\bibinfo {author} {\bibfnamefont {A.-L.}\ \bibnamefont
  {He}}, \bibinfo {author} {\bibfnamefont {L.-R.}\ \bibnamefont {Ding}},
  \bibinfo {author} {\bibfnamefont {Y.}~\bibnamefont {Zhou}}, \bibinfo {author}
  {\bibfnamefont {Y.-F.}\ \bibnamefont {Wang}}, \ and\ \bibinfo {author}
  {\bibfnamefont {C.-D.}\ \bibnamefont {Gong}},\ }\href {\doibase
  10.1103/PhysRevB.100.214109} {\bibfield  {journal} {\bibinfo  {journal}
  {Phys. Rev. B}\ }\textbf {\bibinfo {volume} {100}},\ \bibinfo {pages}
  {214109} (\bibinfo {year} {2019})}\BibitemShut {NoStop}%
\bibitem [{\citenamefont {Benalcazar}\ \emph
  {et~al.}(2017{\natexlab{a}})\citenamefont {Benalcazar}, \citenamefont
  {Bernevig},\ and\ \citenamefont {Hughes}}]{HOTI1}%
  \BibitemOpen
  \bibfield  {author} {\bibinfo {author} {\bibfnamefont {W.~A.}\ \bibnamefont
  {Benalcazar}}, \bibinfo {author} {\bibfnamefont {B.~A.}\ \bibnamefont
  {Bernevig}}, \ and\ \bibinfo {author} {\bibfnamefont {T.~L.}\ \bibnamefont
  {Hughes}},\ }\href {\doibase 10.1126/science.aah6442} {\bibfield  {journal}
  {\bibinfo  {journal} {Science}\ }\textbf {\bibinfo {volume} {357}},\ \bibinfo
  {pages} {61} (\bibinfo {year} {2017}{\natexlab{a}})}\BibitemShut {NoStop}%
\bibitem [{\citenamefont {Benalcazar}\ \emph
  {et~al.}(2017{\natexlab{b}})\citenamefont {Benalcazar}, \citenamefont
  {Bernevig},\ and\ \citenamefont {Hughes}}]{HOTI2}%
  \BibitemOpen
  \bibfield  {author} {\bibinfo {author} {\bibfnamefont {W.~A.}\ \bibnamefont
  {Benalcazar}}, \bibinfo {author} {\bibfnamefont {B.~A.}\ \bibnamefont
  {Bernevig}}, \ and\ \bibinfo {author} {\bibfnamefont {T.~L.}\ \bibnamefont
  {Hughes}},\ }\href {\doibase 10.1103/PhysRevB.96.245115} {\bibfield
  {journal} {\bibinfo  {journal} {Phys. Rev. B}\ }\textbf {\bibinfo {volume}
  {96}},\ \bibinfo {pages} {245115} (\bibinfo {year}
  {2017}{\natexlab{b}})}\BibitemShut {NoStop}%
\bibitem [{\citenamefont {Song}\ \emph {et~al.}(2017)\citenamefont {Song},
  \citenamefont {Fang},\ and\ \citenamefont {Fang}}]{HOTI3}%
  \BibitemOpen
  \bibfield  {author} {\bibinfo {author} {\bibfnamefont {Z.}~\bibnamefont
  {Song}}, \bibinfo {author} {\bibfnamefont {Z.}~\bibnamefont {Fang}}, \ and\
  \bibinfo {author} {\bibfnamefont {C.}~\bibnamefont {Fang}},\ }\href {\doibase
  10.1103/PhysRevLett.119.246402} {\bibfield  {journal} {\bibinfo  {journal}
  {Phys. Rev. Lett.}\ }\textbf {\bibinfo {volume} {119}},\ \bibinfo {pages}
  {246402} (\bibinfo {year} {2017})}\BibitemShut {NoStop}%
\bibitem [{\citenamefont {Langbehn}\ \emph {et~al.}(2017)\citenamefont
  {Langbehn}, \citenamefont {Peng}, \citenamefont {Trifunovic}, \citenamefont
  {von Oppen},\ and\ \citenamefont {Brouwer}}]{HOTI4}%
  \BibitemOpen
  \bibfield  {author} {\bibinfo {author} {\bibfnamefont {J.}~\bibnamefont
  {Langbehn}}, \bibinfo {author} {\bibfnamefont {Y.}~\bibnamefont {Peng}},
  \bibinfo {author} {\bibfnamefont {L.}~\bibnamefont {Trifunovic}}, \bibinfo
  {author} {\bibfnamefont {F.}~\bibnamefont {von Oppen}}, \ and\ \bibinfo
  {author} {\bibfnamefont {P.~W.}\ \bibnamefont {Brouwer}},\ }\href {\doibase
  10.1103/PhysRevLett.119.246401} {\bibfield  {journal} {\bibinfo  {journal}
  {Phys. Rev. Lett.}\ }\textbf {\bibinfo {volume} {119}},\ \bibinfo {pages}
  {246401} (\bibinfo {year} {2017})}\BibitemShut {NoStop}%
\bibitem [{\citenamefont {Schindler}\ \emph {et~al.}(2018)\citenamefont
  {Schindler}, \citenamefont {Cook}, \citenamefont {Vergniory}, \citenamefont
  {Wang}, \citenamefont {Parkin}, \citenamefont {Bernevig},\ and\ \citenamefont
  {Neupert}}]{HOTI5}%
  \BibitemOpen
  \bibfield  {author} {\bibinfo {author} {\bibfnamefont {F.}~\bibnamefont
  {Schindler}}, \bibinfo {author} {\bibfnamefont {A.~M.}\ \bibnamefont {Cook}},
  \bibinfo {author} {\bibfnamefont {M.~G.}\ \bibnamefont {Vergniory}}, \bibinfo
  {author} {\bibfnamefont {Z.}~\bibnamefont {Wang}}, \bibinfo {author}
  {\bibfnamefont {S.~S.~P.}\ \bibnamefont {Parkin}}, \bibinfo {author}
  {\bibfnamefont {B.~A.}\ \bibnamefont {Bernevig}}, \ and\ \bibinfo {author}
  {\bibfnamefont {T.}~\bibnamefont {Neupert}},\ }\href
  {https://advances.sciencemag.org/content/4/6/eaat0346} {\bibfield  {journal}
  {\bibinfo  {journal} {Science Advances}\ }\textbf {\bibinfo {volume} {4}}
  (\bibinfo {year} {2018})}\BibitemShut {NoStop}%
\bibitem [{\citenamefont {Ezawa}(2018{\natexlab{a}})}]{HOTI6}%
  \BibitemOpen
  \bibfield  {author} {\bibinfo {author} {\bibfnamefont {M.}~\bibnamefont
  {Ezawa}},\ }\href {\doibase 10.1103/PhysRevB.98.045125} {\bibfield  {journal}
  {\bibinfo  {journal} {Phys. Rev. B}\ }\textbf {\bibinfo {volume} {98}},\
  \bibinfo {pages} {045125} (\bibinfo {year} {2018}{\natexlab{a}})}\BibitemShut
  {NoStop}%
\bibitem [{\citenamefont {Franca}\ \emph {et~al.}(2018)\citenamefont {Franca},
  \citenamefont {van~den Brink},\ and\ \citenamefont {Fulga}}]{HOTI8}%
  \BibitemOpen
  \bibfield  {author} {\bibinfo {author} {\bibfnamefont {S.}~\bibnamefont
  {Franca}}, \bibinfo {author} {\bibfnamefont {J.}~\bibnamefont {van~den
  Brink}}, \ and\ \bibinfo {author} {\bibfnamefont {I.~C.}\ \bibnamefont
  {Fulga}},\ }\href {\doibase 10.1103/PhysRevB.98.201114} {\bibfield  {journal}
  {\bibinfo  {journal} {Phys. Rev. B}\ }\textbf {\bibinfo {volume} {98}},\
  \bibinfo {pages} {201114} (\bibinfo {year} {2018})}\BibitemShut {NoStop}%
\bibitem [{\citenamefont {Ezawa}(2018{\natexlab{b}})}]{HOTI9}%
  \BibitemOpen
  \bibfield  {author} {\bibinfo {author} {\bibfnamefont {M.}~\bibnamefont
  {Ezawa}},\ }\href {\doibase 10.1103/PhysRevLett.120.026801} {\bibfield
  {journal} {\bibinfo  {journal} {Phys. Rev. Lett.}\ }\textbf {\bibinfo
  {volume} {120}},\ \bibinfo {pages} {026801} (\bibinfo {year}
  {2018}{\natexlab{b}})}\BibitemShut {NoStop}%
\bibitem [{\citenamefont {Trifunovic}\ and\ \citenamefont
  {Brouwer}(2019)}]{HOTI10}%
  \BibitemOpen
  \bibfield  {author} {\bibinfo {author} {\bibfnamefont {L.}~\bibnamefont
  {Trifunovic}}\ and\ \bibinfo {author} {\bibfnamefont {P.~W.}\ \bibnamefont
  {Brouwer}},\ }\href {\doibase 10.1103/PhysRevX.9.011012} {\bibfield
  {journal} {\bibinfo  {journal} {Phys. Rev. X}\ }\textbf {\bibinfo {volume}
  {9}},\ \bibinfo {pages} {011012} (\bibinfo {year} {2019})}\BibitemShut
  {NoStop}%
\bibitem [{\citenamefont {Chen}\ \emph
  {et~al.}(2020{\natexlab{a}})\citenamefont {Chen}, \citenamefont {Chen},
  \citenamefont {Gao}, \citenamefont {Zhou},\ and\ \citenamefont
  {Xu}}]{HOTI11}%
  \BibitemOpen
  \bibfield  {author} {\bibinfo {author} {\bibfnamefont {R.}~\bibnamefont
  {Chen}}, \bibinfo {author} {\bibfnamefont {C.-Z.}\ \bibnamefont {Chen}},
  \bibinfo {author} {\bibfnamefont {J.-H.}\ \bibnamefont {Gao}}, \bibinfo
  {author} {\bibfnamefont {B.}~\bibnamefont {Zhou}}, \ and\ \bibinfo {author}
  {\bibfnamefont {D.-H.}\ \bibnamefont {Xu}},\ }\href {\doibase
  10.1103/PhysRevLett.124.036803} {\bibfield  {journal} {\bibinfo  {journal}
  {Phys. Rev. Lett.}\ }\textbf {\bibinfo {volume} {124}},\ \bibinfo {pages}
  {036803} (\bibinfo {year} {2020}{\natexlab{a}})}\BibitemShut {NoStop}%
\bibitem [{\citenamefont {Lee}\ \emph {et~al.}(2020)\citenamefont {Lee},
  \citenamefont {Furusaki},\ and\ \citenamefont {Yang}}]{HOTI12}%
  \BibitemOpen
  \bibfield  {author} {\bibinfo {author} {\bibfnamefont {E.}~\bibnamefont
  {Lee}}, \bibinfo {author} {\bibfnamefont {A.}~\bibnamefont {Furusaki}}, \
  and\ \bibinfo {author} {\bibfnamefont {B.-J.}\ \bibnamefont {Yang}},\ }\href
  {\doibase 10.1103/PhysRevB.101.241109} {\bibfield  {journal} {\bibinfo
  {journal} {Phys. Rev. B}\ }\textbf {\bibinfo {volume} {101}},\ \bibinfo
  {pages} {241109} (\bibinfo {year} {2020})}\BibitemShut {NoStop}%
\bibitem [{\citenamefont {He}\ \emph {et~al.}()\citenamefont {He},
  \citenamefont {Wang},\ and\ \citenamefont {Yao}}]{HOTI13}%
  \BibitemOpen
  \bibfield  {author} {\bibinfo {author} {\bibfnamefont {A.-L.}\ \bibnamefont
  {He}}, \bibinfo {author} {\bibfnamefont {Y.-F.}\ \bibnamefont {Wang}}, \ and\
  \bibinfo {author} {\bibfnamefont {H.}~\bibnamefont {Yao}},\ }\href@noop {} {\
  }\bibinfo {note} {(unpublished)}\BibitemShut {NoStop}%
\bibitem [{\citenamefont {Lau}\ \emph {et~al.}(2015)\citenamefont {Lau},
  \citenamefont {Ortix},\ and\ \citenamefont {van~den Brink}}]{SSH4}%
  \BibitemOpen
  \bibfield  {author} {\bibinfo {author} {\bibfnamefont {A.}~\bibnamefont
  {Lau}}, \bibinfo {author} {\bibfnamefont {C.}~\bibnamefont {Ortix}}, \ and\
  \bibinfo {author} {\bibfnamefont {J.}~\bibnamefont {van~den Brink}},\ }\href
  {\doibase 10.1103/PhysRevLett.115.216805} {\bibfield  {journal} {\bibinfo
  {journal} {Phys. Rev. Lett.}\ }\textbf {\bibinfo {volume} {115}},\ \bibinfo
  {pages} {216805} (\bibinfo {year} {2015})}\BibitemShut {NoStop}%
\bibitem [{\citenamefont {Liu}\ and\ \citenamefont {Wakabayashi}(2017)}]{Liu1}%
  \BibitemOpen
  \bibfield  {author} {\bibinfo {author} {\bibfnamefont {F.}~\bibnamefont
  {Liu}}\ and\ \bibinfo {author} {\bibfnamefont {K.}~\bibnamefont
  {Wakabayashi}},\ }\href {\doibase 10.1103/PhysRevLett.118.076803} {\bibfield
  {journal} {\bibinfo  {journal} {Phys. Rev. Lett.}\ }\textbf {\bibinfo
  {volume} {118}},\ \bibinfo {pages} {076803} (\bibinfo {year}
  {2017})}\BibitemShut {NoStop}%
\bibitem [{\citenamefont {Thouless}\ \emph {et~al.}(1982)\citenamefont
  {Thouless}, \citenamefont {Kohmoto}, \citenamefont {Nightingale},\ and\
  \citenamefont {den Nijs}}]{TKNN}%
  \BibitemOpen
  \bibfield  {author} {\bibinfo {author} {\bibfnamefont {D.~J.}\ \bibnamefont
  {Thouless}}, \bibinfo {author} {\bibfnamefont {M.}~\bibnamefont {Kohmoto}},
  \bibinfo {author} {\bibfnamefont {M.~P.}\ \bibnamefont {Nightingale}}, \ and\
  \bibinfo {author} {\bibfnamefont {M.}~\bibnamefont {den Nijs}},\ }\href
  {\doibase 10.1103/PhysRevLett.49.405} {\bibfield  {journal} {\bibinfo
  {journal} {Phys. Rev. Lett.}\ }\textbf {\bibinfo {volume} {49}},\ \bibinfo
  {pages} {405} (\bibinfo {year} {1982})}\BibitemShut {NoStop}%
\bibitem [{sup()}]{suppl}%
  \BibitemOpen
  \href@noop {} {\ }\bibinfo {note} {See Supplemental Material for more details
  about the topological invariants, characterization of these topological
  states, and the robustness of the topological states, etc.}\BibitemShut
  {Stop}%
\bibitem [{\citenamefont {Kitaev}(2006)}]{LCHN1}%
  \BibitemOpen
  \bibfield  {author} {\bibinfo {author} {\bibfnamefont {A.}~\bibnamefont
  {Kitaev}},\ }\href {\doibase https://doi.org/10.1016/j.aop.2005.10.005}
  {\bibfield  {journal} {\bibinfo  {journal} {Annals of Physics}\ }\textbf
  {\bibinfo {volume} {321}},\ \bibinfo {pages} {2 } (\bibinfo {year}
  {2006})}\BibitemShut {NoStop}%
\bibitem [{\citenamefont {Cheng}\ \emph {et~al.}(2018)\citenamefont {Cheng},
  \citenamefont {Luo}, \citenamefont {He},\ and\ \citenamefont {Wang}}]{HC2}%
  \BibitemOpen
  \bibfield  {author} {\bibinfo {author} {\bibfnamefont {Q.~Q.}\ \bibnamefont
  {Cheng}}, \bibinfo {author} {\bibfnamefont {W.~W.}\ \bibnamefont {Luo}},
  \bibinfo {author} {\bibfnamefont {A.~L.}\ \bibnamefont {He}}, \ and\ \bibinfo
  {author} {\bibfnamefont {Y.~F.}\ \bibnamefont {Wang}},\ }\href {\doibase
  10.1088/1361-648X/aad51f} {\bibfield  {journal} {\bibinfo  {journal} {J Phys
  Condens Matter}\ }\textbf {\bibinfo {volume} {30}},\ \bibinfo {pages}
  {355502} (\bibinfo {year} {2018})}\BibitemShut {NoStop}%
\bibitem [{\citenamefont {Hughes}\ \emph {et~al.}(2011)\citenamefont {Hughes},
  \citenamefont {Prodan},\ and\ \citenamefont {Bernevig}}]{ISTIs}%
  \BibitemOpen
  \bibfield  {author} {\bibinfo {author} {\bibfnamefont {T.~L.}\ \bibnamefont
  {Hughes}}, \bibinfo {author} {\bibfnamefont {E.}~\bibnamefont {Prodan}}, \
  and\ \bibinfo {author} {\bibfnamefont {B.~A.}\ \bibnamefont {Bernevig}},\
  }\href {\doibase 10.1103/PhysRevB.83.245132} {\bibfield  {journal} {\bibinfo
  {journal} {Phys. Rev. B}\ }\textbf {\bibinfo {volume} {83}},\ \bibinfo
  {pages} {245132} (\bibinfo {year} {2011})}\BibitemShut {NoStop}%
\bibitem [{\citenamefont {Fang}\ \emph {et~al.}(2012)\citenamefont {Fang},
  \citenamefont {Gilbert},\ and\ \citenamefont {Bernevig}}]{CFang}%
  \BibitemOpen
  \bibfield  {author} {\bibinfo {author} {\bibfnamefont {C.}~\bibnamefont
  {Fang}}, \bibinfo {author} {\bibfnamefont {M.~J.}\ \bibnamefont {Gilbert}}, \
  and\ \bibinfo {author} {\bibfnamefont {B.~A.}\ \bibnamefont {Bernevig}},\
  }\href {\doibase 10.1103/PhysRevB.86.115112} {\bibfield  {journal} {\bibinfo
  {journal} {Phys. Rev. B}\ }\textbf {\bibinfo {volume} {86}},\ \bibinfo
  {pages} {115112} (\bibinfo {year} {2012})}\BibitemShut {NoStop}%
\bibitem [{\citenamefont {Liu}\ \emph {et~al.}(2019)\citenamefont {Liu},
  \citenamefont {Deng},\ and\ \citenamefont {Wakabayashi}}]{polarcharge4}%
  \BibitemOpen
  \bibfield  {author} {\bibinfo {author} {\bibfnamefont {F.}~\bibnamefont
  {Liu}}, \bibinfo {author} {\bibfnamefont {H.-Y.}\ \bibnamefont {Deng}}, \
  and\ \bibinfo {author} {\bibfnamefont {K.}~\bibnamefont {Wakabayashi}},\
  }\href {\doibase 10.1103/PhysRevLett.122.086804} {\bibfield  {journal}
  {\bibinfo  {journal} {Phys. Rev. Lett.}\ }\textbf {\bibinfo {volume} {122}},\
  \bibinfo {pages} {086804} (\bibinfo {year} {2019})}\BibitemShut {NoStop}%
\bibitem [{\citenamefont {Aidelsburger}\ \emph {et~al.}(2013)\citenamefont
  {Aidelsburger}, \citenamefont {Atala}, \citenamefont {Lohse}, \citenamefont
  {Barreiro}, \citenamefont {Paredes},\ and\ \citenamefont {Bloch}}]{Hofs}%
  \BibitemOpen
  \bibfield  {author} {\bibinfo {author} {\bibfnamefont {M.}~\bibnamefont
  {Aidelsburger}}, \bibinfo {author} {\bibfnamefont {M.}~\bibnamefont {Atala}},
  \bibinfo {author} {\bibfnamefont {M.}~\bibnamefont {Lohse}}, \bibinfo
  {author} {\bibfnamefont {J.~T.}\ \bibnamefont {Barreiro}}, \bibinfo {author}
  {\bibfnamefont {B.}~\bibnamefont {Paredes}}, \ and\ \bibinfo {author}
  {\bibfnamefont {I.}~\bibnamefont {Bloch}},\ }\href {\doibase
  10.1103/PhysRevLett.111.185301} {\bibfield  {journal} {\bibinfo  {journal}
  {Phys. Rev. Lett.}\ }\textbf {\bibinfo {volume} {111}},\ \bibinfo {pages}
  {185301} (\bibinfo {year} {2013})}\BibitemShut {NoStop}%
\bibitem [{\citenamefont {Serra-Garcia}\ \emph {et~al.}(2018)\citenamefont
  {Serra-Garcia}, \citenamefont {Peri}, \citenamefont {Süsstrunk},
  \citenamefont {Bilal}, \citenamefont {Larsen}, \citenamefont {Villanueva},\
  and\ \citenamefont {Huber}}]{Phot_HOTI}%
  \BibitemOpen
  \bibfield  {author} {\bibinfo {author} {\bibfnamefont {M.}~\bibnamefont
  {Serra-Garcia}}, \bibinfo {author} {\bibfnamefont {V.}~\bibnamefont {Peri}},
  \bibinfo {author} {\bibfnamefont {R.}~\bibnamefont {Süsstrunk}}, \bibinfo
  {author} {\bibfnamefont {O.~R.}\ \bibnamefont {Bilal}}, \bibinfo {author}
  {\bibfnamefont {T.}~\bibnamefont {Larsen}}, \bibinfo {author} {\bibfnamefont
  {L.~G.}\ \bibnamefont {Villanueva}}, \ and\ \bibinfo {author} {\bibfnamefont
  {S.~D.}\ \bibnamefont {Huber}},\ }\href {\doibase 10.1038/nature25156}
  {\bibfield  {journal} {\bibinfo  {journal} {Nature}\ }\textbf {\bibinfo
  {volume} {555}},\ \bibinfo {pages} {342} (\bibinfo {year}
  {2018})}\BibitemShut {NoStop}%
\bibitem [{\citenamefont {Peterson}\ \emph {et~al.}(2018)\citenamefont
  {Peterson}, \citenamefont {Benalcazar}, \citenamefont {Hughes},\ and\
  \citenamefont {Bahl}}]{Phot_HI}%
  \BibitemOpen
  \bibfield  {author} {\bibinfo {author} {\bibfnamefont {C.~W.}\ \bibnamefont
  {Peterson}}, \bibinfo {author} {\bibfnamefont {W.~A.}\ \bibnamefont
  {Benalcazar}}, \bibinfo {author} {\bibfnamefont {T.~L.}\ \bibnamefont
  {Hughes}}, \ and\ \bibinfo {author} {\bibfnamefont {G.}~\bibnamefont
  {Bahl}},\ }\href {\doibase 10.1038/nature25777} {\bibfield  {journal}
  {\bibinfo  {journal} {Nature}\ }\textbf {\bibinfo {volume} {555}},\ \bibinfo
  {pages} {346} (\bibinfo {year} {2018})}\BibitemShut {NoStop}%
\bibitem [{\citenamefont {Imhof}\ \emph {et~al.}(2018)\citenamefont {Imhof},
  \citenamefont {Berger}, \citenamefont {Bayer}, \citenamefont {Brehm},
  \citenamefont {Molenkamp}, \citenamefont {Kiessling}, \citenamefont
  {Schindler}, \citenamefont {Lee}, \citenamefont {Greiter}, \citenamefont
  {Neupert},\ and\ \citenamefont {Thomale}}]{Circ_HOTI}%
  \BibitemOpen
  \bibfield  {author} {\bibinfo {author} {\bibfnamefont {S.}~\bibnamefont
  {Imhof}}, \bibinfo {author} {\bibfnamefont {C.}~\bibnamefont {Berger}},
  \bibinfo {author} {\bibfnamefont {F.}~\bibnamefont {Bayer}}, \bibinfo
  {author} {\bibfnamefont {J.}~\bibnamefont {Brehm}}, \bibinfo {author}
  {\bibfnamefont {L.~W.}\ \bibnamefont {Molenkamp}}, \bibinfo {author}
  {\bibfnamefont {T.}~\bibnamefont {Kiessling}}, \bibinfo {author}
  {\bibfnamefont {F.}~\bibnamefont {Schindler}}, \bibinfo {author}
  {\bibfnamefont {C.~H.}\ \bibnamefont {Lee}}, \bibinfo {author} {\bibfnamefont
  {M.}~\bibnamefont {Greiter}}, \bibinfo {author} {\bibfnamefont
  {T.}~\bibnamefont {Neupert}}, \ and\ \bibinfo {author} {\bibfnamefont
  {R.}~\bibnamefont {Thomale}},\ }\href {\doibase 10.1038/s41567-018-0246-1}
  {\bibfield  {journal} {\bibinfo  {journal} {Nature Physics}\ }\textbf
  {\bibinfo {volume} {14}},\ \bibinfo {pages} {925} (\bibinfo {year}
  {2018})}\BibitemShut {NoStop}%
\bibitem [{\citenamefont {Peterson}\ \emph {et~al.}(2020)\citenamefont
  {Peterson}, \citenamefont {Li}, \citenamefont {Benalcazar}, \citenamefont
  {Hughes},\ and\ \citenamefont {Bahl}}]{Peterson}%
  \BibitemOpen
  \bibfield  {author} {\bibinfo {author} {\bibfnamefont {C.~W.}\ \bibnamefont
  {Peterson}}, \bibinfo {author} {\bibfnamefont {T.}~\bibnamefont {Li}},
  \bibinfo {author} {\bibfnamefont {W.~A.}\ \bibnamefont {Benalcazar}},
  \bibinfo {author} {\bibfnamefont {T.~L.}\ \bibnamefont {Hughes}}, \ and\
  \bibinfo {author} {\bibfnamefont {G.}~\bibnamefont {Bahl}},\ }\href {\doibase
  10.1126/science.aba7604} {\bibfield  {journal} {\bibinfo  {journal}
  {Science}\ }\textbf {\bibinfo {volume} {368}},\ \bibinfo {pages} {1114}
  (\bibinfo {year} {2020})}\BibitemShut {NoStop}%
\bibitem [{\citenamefont {Hofmann}\ \emph {et~al.}(2019)\citenamefont
  {Hofmann}, \citenamefont {Helbig}, \citenamefont {Lee}, \citenamefont
  {Greiter},\ and\ \citenamefont {Thomale}}]{Circ_chern}%
  \BibitemOpen
  \bibfield  {author} {\bibinfo {author} {\bibfnamefont {T.}~\bibnamefont
  {Hofmann}}, \bibinfo {author} {\bibfnamefont {T.}~\bibnamefont {Helbig}},
  \bibinfo {author} {\bibfnamefont {C.~H.}\ \bibnamefont {Lee}}, \bibinfo
  {author} {\bibfnamefont {M.}~\bibnamefont {Greiter}}, \ and\ \bibinfo
  {author} {\bibfnamefont {R.}~\bibnamefont {Thomale}},\ }\href {\doibase
  10.1103/PhysRevLett.122.247702} {\bibfield  {journal} {\bibinfo  {journal}
  {Phys. Rev. Lett.}\ }\textbf {\bibinfo {volume} {122}},\ \bibinfo {pages}
  {247702} (\bibinfo {year} {2019})}\BibitemShut {NoStop}%
\bibitem [{\citenamefont {Chen}\ \emph
  {et~al.}(2020{\natexlab{b}})\citenamefont {Chen}, \citenamefont {Sharpe},
  \citenamefont {Fox}, \citenamefont {Zhang}, \citenamefont {Wang},
  \citenamefont {Jiang}, \citenamefont {Lyu}, \citenamefont {Li}, \citenamefont
  {Watanabe}, \citenamefont {Taniguchi}, \citenamefont {Shi}, \citenamefont
  {Senthil}, \citenamefont {Goldhaber-Gordon}, \citenamefont {Zhang},\ and\
  \citenamefont {Wang}}]{TGM}%
  \BibitemOpen
  \bibfield  {author} {\bibinfo {author} {\bibfnamefont {G.}~\bibnamefont
  {Chen}}, \bibinfo {author} {\bibfnamefont {A.~L.}\ \bibnamefont {Sharpe}},
  \bibinfo {author} {\bibfnamefont {E.~J.}\ \bibnamefont {Fox}}, \bibinfo
  {author} {\bibfnamefont {Y.~H.}\ \bibnamefont {Zhang}}, \bibinfo {author}
  {\bibfnamefont {S.}~\bibnamefont {Wang}}, \bibinfo {author} {\bibfnamefont
  {L.}~\bibnamefont {Jiang}}, \bibinfo {author} {\bibfnamefont
  {B.}~\bibnamefont {Lyu}}, \bibinfo {author} {\bibfnamefont {H.}~\bibnamefont
  {Li}}, \bibinfo {author} {\bibfnamefont {K.}~\bibnamefont {Watanabe}},
  \bibinfo {author} {\bibfnamefont {T.}~\bibnamefont {Taniguchi}}, \bibinfo
  {author} {\bibfnamefont {Z.}~\bibnamefont {Shi}}, \bibinfo {author}
  {\bibfnamefont {T.}~\bibnamefont {Senthil}}, \bibinfo {author} {\bibfnamefont
  {D.}~\bibnamefont {Goldhaber-Gordon}}, \bibinfo {author} {\bibfnamefont
  {Y.}~\bibnamefont {Zhang}}, \ and\ \bibinfo {author} {\bibfnamefont
  {F.}~\bibnamefont {Wang}},\ }\href {\doibase 10.1038/s41586-020-2049-7}
  {\bibfield  {journal} {\bibinfo  {journal} {Nature}\ }\textbf {\bibinfo
  {volume} {579}},\ \bibinfo {pages} {56} (\bibinfo {year}
  {2020}{\natexlab{b}})}\BibitemShut {NoStop}%
\bibitem [{\citenamefont {Serlin}\ \emph {et~al.}(2020)\citenamefont {Serlin},
  \citenamefont {Tschirhart}, \citenamefont {Polshyn}, \citenamefont {Zhang},
  \citenamefont {Zhu}, \citenamefont {Watanabe}, \citenamefont {Taniguchi},
  \citenamefont {Balents},\ and\ \citenamefont {Young}}]{TBG_MH}%
  \BibitemOpen
  \bibfield  {author} {\bibinfo {author} {\bibfnamefont {M.}~\bibnamefont
  {Serlin}}, \bibinfo {author} {\bibfnamefont {C.~L.}\ \bibnamefont
  {Tschirhart}}, \bibinfo {author} {\bibfnamefont {H.}~\bibnamefont {Polshyn}},
  \bibinfo {author} {\bibfnamefont {Y.}~\bibnamefont {Zhang}}, \bibinfo
  {author} {\bibfnamefont {J.}~\bibnamefont {Zhu}}, \bibinfo {author}
  {\bibfnamefont {K.}~\bibnamefont {Watanabe}}, \bibinfo {author}
  {\bibfnamefont {T.}~\bibnamefont {Taniguchi}}, \bibinfo {author}
  {\bibfnamefont {L.}~\bibnamefont {Balents}}, \ and\ \bibinfo {author}
  {\bibfnamefont {A.~F.}\ \bibnamefont {Young}},\ }\href {\doibase
  10.1126/science.aay5533} {\bibfield  {journal} {\bibinfo  {journal}
  {Science}\ }\textbf {\bibinfo {volume} {367}},\ \bibinfo {pages} {900}
  (\bibinfo {year} {2020})}\BibitemShut {NoStop}%
\bibitem [{\citenamefont {Nuckolls}\ \emph {et~al.}(2020)\citenamefont
  {Nuckolls}, \citenamefont {Oh}, \citenamefont {Wong}, \citenamefont {Lian},
  \citenamefont {Watanabe}, \citenamefont {Taniguchi}, \citenamefont
  {Bernevig},\ and\ \citenamefont {Yazdani}}]{SC_TBG}%
  \BibitemOpen
  \bibfield  {author} {\bibinfo {author} {\bibfnamefont {K.~P.}\ \bibnamefont
  {Nuckolls}}, \bibinfo {author} {\bibfnamefont {M.}~\bibnamefont {Oh}},
  \bibinfo {author} {\bibfnamefont {D.}~\bibnamefont {Wong}}, \bibinfo {author}
  {\bibfnamefont {B.}~\bibnamefont {Lian}}, \bibinfo {author} {\bibfnamefont
  {K.}~\bibnamefont {Watanabe}}, \bibinfo {author} {\bibfnamefont
  {T.}~\bibnamefont {Taniguchi}}, \bibinfo {author} {\bibfnamefont {B.~A.}\
  \bibnamefont {Bernevig}}, \ and\ \bibinfo {author} {\bibfnamefont
  {A.}~\bibnamefont {Yazdani}},\ }\href {\doibase 10.1038/s41586-020-3028-8}
  {\bibfield  {journal} {\bibinfo  {journal} {Nature}\ }\textbf {\bibinfo
  {volume} {588}},\ \bibinfo {pages} {610} (\bibinfo {year}
  {2020})}\BibitemShut {NoStop}%
\end{thebibliography}%

\widetext
\appendix

\section*{SUPPLEMENTARY MATERIALs FOR ``Topological states in a dimerized system with staggered magnetic fluxes''}
\setcounter{figure}{0}
\setcounter{equation}{0}
\renewcommand \thefigure{S\arabic{figure}}
\renewcommand \theequation{S\arabic{equation}}
\setcounter{section}{0}
\renewcommand \thesection{S\arabic{section}}

In main text, we investigate two kinds of topological insulator states in the extended $2$D SSH model. At half-filling, we find a ${\cal C}=2$ Chern insulating (CI) state. With open boundary condition, the in-gap states are divided into two parts---one diffuses into the near-edge and the other gathers around the corner, yielding the isolated corner states but without fractional corner charge and the metallic near-edge states. Those special boundary states significantly differ from the conventional CIs, which host the well-edge localized boundary states. We identify the present CI by the Wannier bands, and find the shift of the Wannier bands. Moreover, we find another topological state with zero Chern number at quarter-filling. In this topological state, the boundary states bypass the corner and localize robustly along the edges, yielding the absence of the corner states. In the supplement, we show more details of the topological invariants, characterization of our present topological states, and the robustness of the topological states, etc.

\section{Topological invariants} \label{Topindex}
Topological invariant is an important index to characterize the topological properties of system. Here, we show details of topological invariants used in the main text, including the Chern number, the real-space Chern number, Wannier centers and $2$D Zak phase.

The Chern number ${\cal C}_n$ can be calculated by integrating the Berry curvature ${\cal F}_{n}({\bf k})$ over the first Brillouin zone~\cite{TKNN,KG2},
\begin{equation}\label{Ch_num}
{\cal C}_n = \frac{1}{2\pi}\int_{BZ}d^2{\bf k}{\cal  F}_{n}({\bf k}) .
\end{equation}
Here, ${\cal F}_{n}({\bf k}) = \nabla\times{\cal A}_n({\bf k})$ with the berry connection ${\cal A}_n({\bf k})=-i\langle u_{n\bf{k}}\vert\nabla_{\bf{k}}\vert u_{n\bf{k}}\rangle$  and $|u_{n\bf{k}}\rangle$ is Bloch wave functions of the $n$th band. Chern number can be expressed in inversion systems at high symmetric points~\cite{ISTIs,CFang} as,
\begin{equation}\label{Chern_syp}
(-1)^{\cal C} = \prod^n_{i} \zeta_i(\Gamma) \zeta_i(X) \zeta_i(Y) \zeta_i(M),
\end{equation}
where $\zeta$ indicates the parity, the eigenvalue of inversion operator $\cal{P}$, $n$ is the number of occupied bands.

\begin{figure}[!htb]
\includegraphics[scale=0.9]{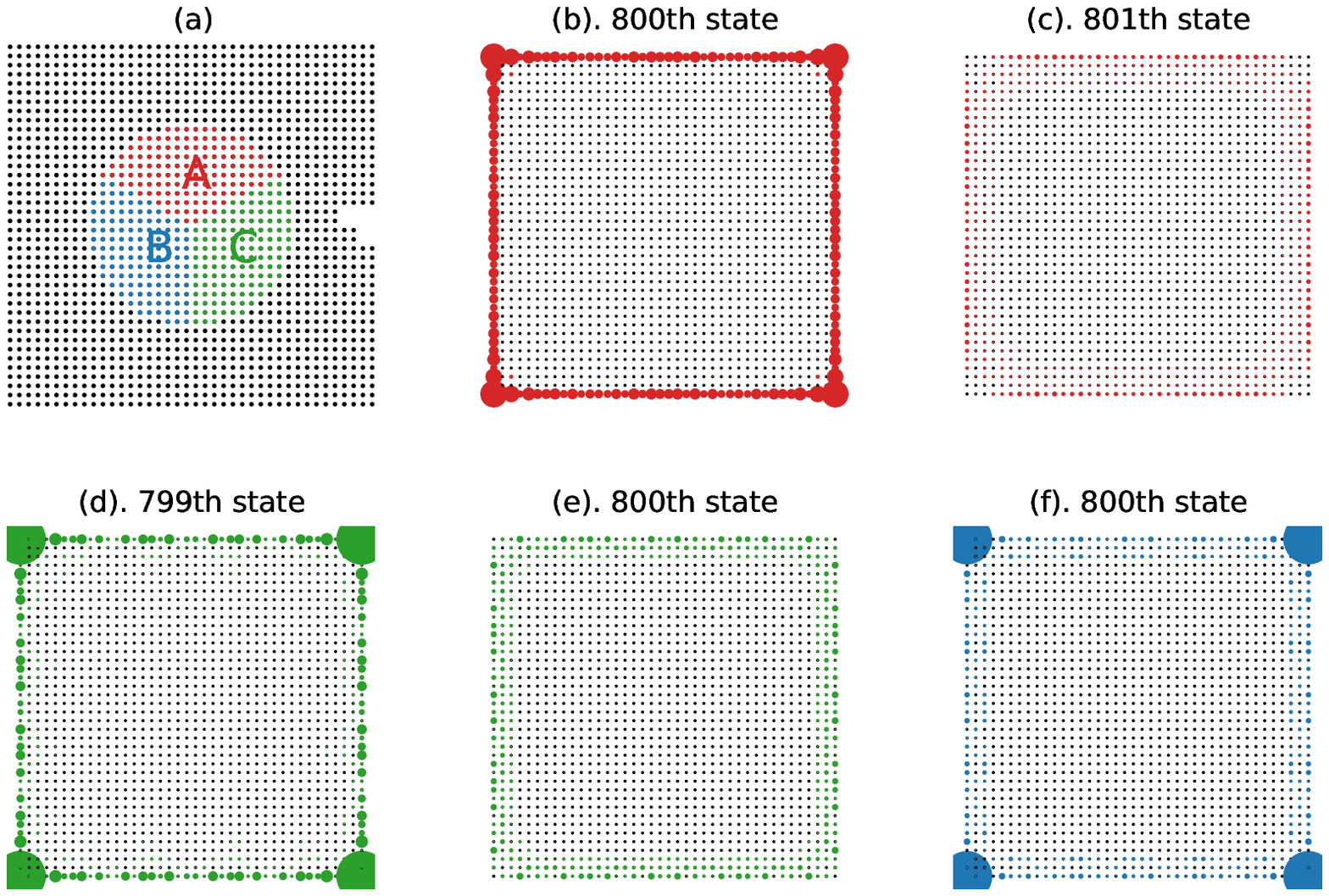}
\caption{(color online). Evolution of the boundary states with various $v$. (a) Schematic scheme to calculate the real-space Chern number by Kitaev formula by dividing the bulk into three distinct neighboring regions marked with ``A", ``B" and ``C". (b)-(f) the spatial distribution of the in-gap states in the $\mathcal{C}=2$ CI states. $v=1.0$ for (b) and (c), $v=1.5$ for(d) and (e), and $v=2.0$ for (f). Here, $w^{\prime}=0.75$ and $\phi=-0.25\pi$}
\label{ACI_x}
\end{figure}

The topological characterization of CIs without translational symmetry can be further identified by the real-space Chern number. In the main text, we adopt Kitaev formula to calculate real-space Chern number, i.e.,
\begin{equation}\label{kit_for}
{\cal C}_{R} = 12\pi i \sum_{j\in A} \sum_{k\in B} \sum_{l\in C} (P_{jk}P_{kl}P_{lj} - P_{jl}P_{lk}P_{kj}).
\end{equation}
A circle region is chosen in bulk and cut into three distinct neighboring regions arranged in a counterclockwise order, marked with ``A", ``B" and ``C" [Fig.~\ref{ACI_x} (a)]. $j$, $k$ and $l$ mark the sites in A, B, and C regions, respectively~\cite{LCHN1}. $\hat{P}$ is the projection operator defined up to Fermi energy $E_F$ and $\hat{P}=\sum_{E_n<E_F} \vert\phi_n\rangle \langle \phi_n \vert$ with the matrix elements $P_{ij}=\sum_{E_n<E_F} \phi_n(r_j) \phi^{*}_n(r_i)$.

We compute the Wannier centers of the present CI states in the main text. Wannier centers can be implemented with the aid of the Wilson loop~\cite{HOTI1,HOTI2}. In the thermodynamic limit, Wannier bands along $k_y$ are obtained based on the Wilson loop operator~\cite{HOTI1,HOTI2},
\begin{equation}\label{Wil_loop}
{\cal W}_{(k_x+2\pi,k_y)\leftarrow(k_x,k_y)} \equiv {\cal W}_{x,{\bf k}} = \lim\limits_{n \to \infty} F_{n-1}...F_{1}F_{0}.
\end{equation}
Here, $[F_i]_{mn}= \langle u_m(k_{i+1},k_y) \vert u_n(k_{i},k_y)\rangle$, $\vert u_n(k_x,k_y)\rangle$ is the $n$-th eigen wave-function, $k_i=\frac{2\pi}{n}i$ and $m,n=1,2...N_{occ}$. By diagonalizing these Wilson loop operators, we obtain the Wannier spectra on the torus geometry. Since the Wilson loop operator is unitary~\cite{HOTI1,HOTI2,HOTI8}, the eigenvalue equation can be written as,
\begin{equation}\label{Wan_band_eig}
{\cal W}_{x,{\bf k}} |\nu^{j}_{x,{\bf k}}\rangle = {\rm exp}(2\pi i \nu^{j}_{x}(k_y))|\nu^{j}_{x,{\bf k}}\rangle,
\end{equation}
where $\nu^{j}_{x}(k_y)$ corresponds to the Wannier center of the $j$th Wannier function and determines the Wannier bands.

The zero-Chern number topological state in dimerized lattice with inversion symmetry is characterized by the $2$D Zak phase or wave polarization, which is expressed based on the parity as~\cite{CFang,Liu1},
\begin{equation}\label{Zak}
p_i = \sum_n p^n_i,  \qquad   p^n_i = 1/2(q^n_i {\rm mod}  2)    ,  \qquad   (-1)^{q^n_i} = \frac{\zeta(X_i)}{\zeta(\Gamma)},
\end{equation}
where $i$ denotes $x-$ or $y-$direction.

\begin{figure}[!htb]
\includegraphics[scale=0.9]{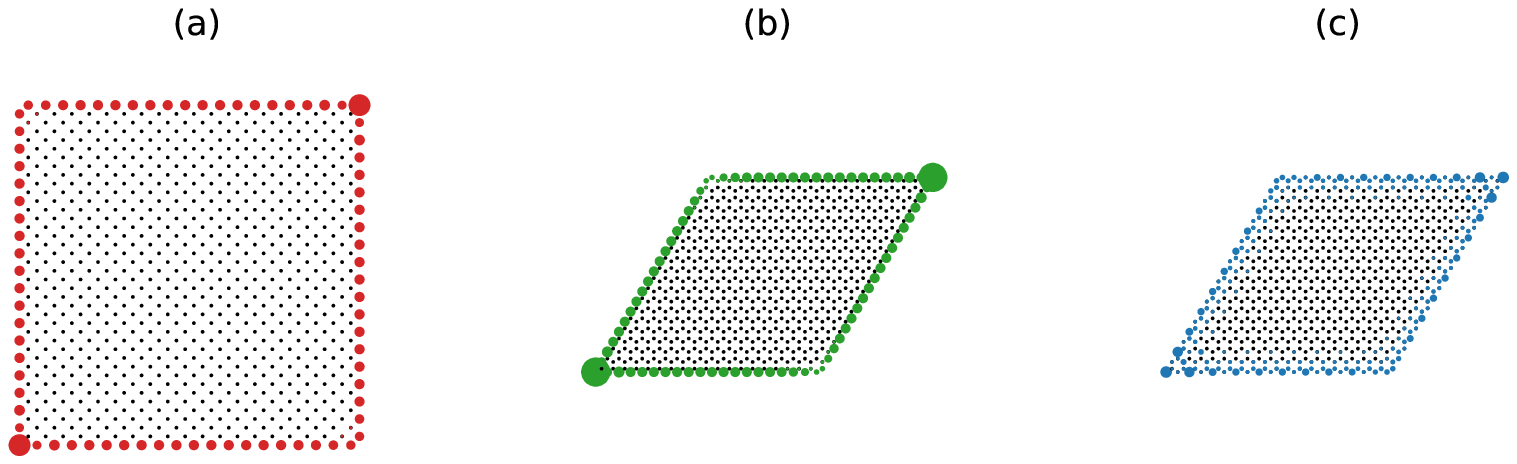}
\caption{(color online). Edge states of conventional CIs. We present the robust edge states of (a) Checkerboard model, (b) Haldane model with Chern number $\mathcal{C}=1$ and (c) Haldane model with Chern number $\mathcal{C}=2$.}
\label{convCI}
\end{figure}

\section{ ${\cal C}=2$ CI states} \label{ATIS0}
For a weak dimerized potential ($v=1.0$ and $v=1.5$), some of boundary states are robustly localized along the edges [Fig.~\ref{ACI_x} (b) and (d)], in analogy to the conventional CIs. However, other boundary states bypass the corners and slightly extend into the near-edges [Fig.~\ref{ACI_x} (c) and (e)], which is in contrast to the conventional CIs (Fig.~\ref{convCI}). When the boundary states robustly localized along the edges [Fig.~\ref{ACI_x} (b) and (d)], a weak but visible portion gathers at corners even at weak $v$. The corner states enhances with $v$ [Fig.~\ref{ACI_x} (f) at $v$=2.0]. When $v>2.2$, the isolated corner states and the near-edge states are evident.

\begin{figure}[!htb]
\includegraphics[scale=0.7]{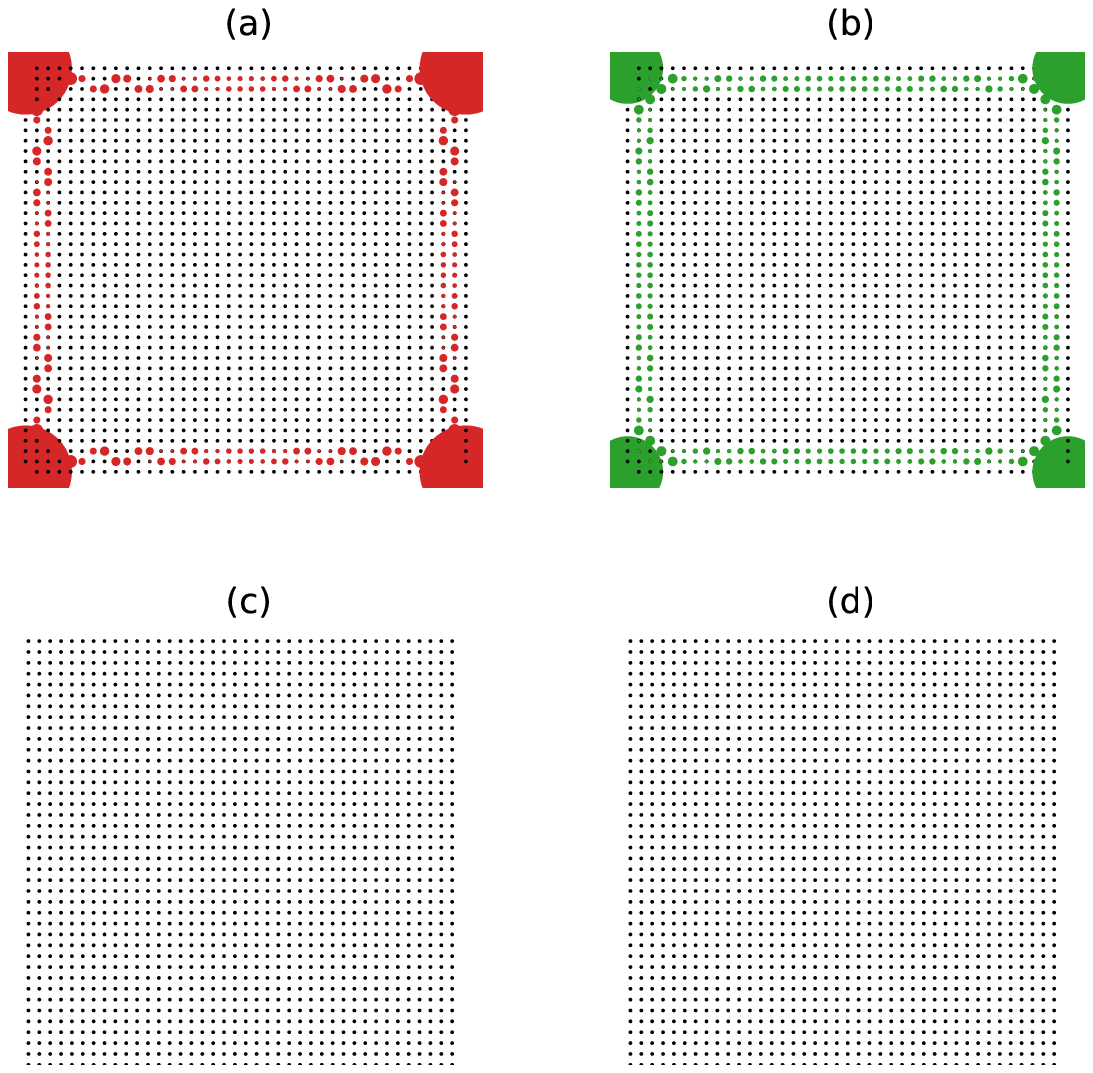}
\caption{(color online). The distributions of the anomalous boundary at half-filling in the strong dimerization limit.(a) $v=20.0$ and (b) $v=50.0$. There is no excessive charge around the corners in (c) and (d) when $v=20.0$ and $v=50.0$.  Here, $w^{\prime}=0.75$ and $\phi=-0.25\pi$}
\label{ACI1}
\end{figure}

\begin{figure}[!htb]
\includegraphics[scale=0.8]{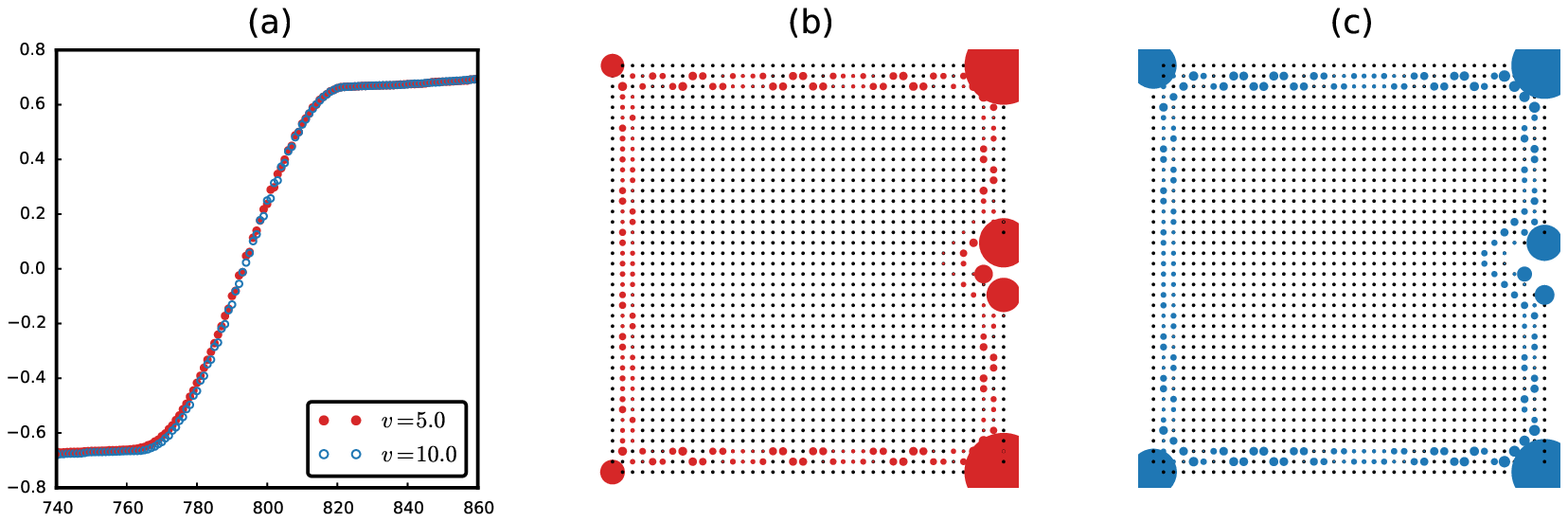}
\caption{(color online). Eigen-energies and distribution of boundary state in real-space. Here, we choose a $20\times20$ square lattice with a defect. (a) Eigen-energies $E$ versus the state index with $v=5.0$ and $10.0$. And we plot the distributions of boundary states with $v=5.0$ in (b) and $v=10.0$ in (c). The hopping and flux parameters are chosen as $w^{\prime}=0.75$ and $\phi=-0.25\pi$. }
\label{ACI2}
\end{figure}

For strong dimerization, the boundary states gather at corners and diffuse into quasi-edge, even in the strong dimerization limit. We show the spatial distribution of the $800th$ state (edge state) in a $1600$-site square lattice with $v=20.0$ and $v=50.0$ in Fig.~\ref{ACI1} (a) and (b). Evidently, the above mentioned features of the boundary states remain robust. The corner charge of the present CI state is further calculated, and no fractional corner charge is observed. Hence, this provides a proof to identify the topological state as a CI state, instead of the higher-order topological insulator (HOTI) state.

\begin{figure}[!htb]
\includegraphics[scale=0.9]{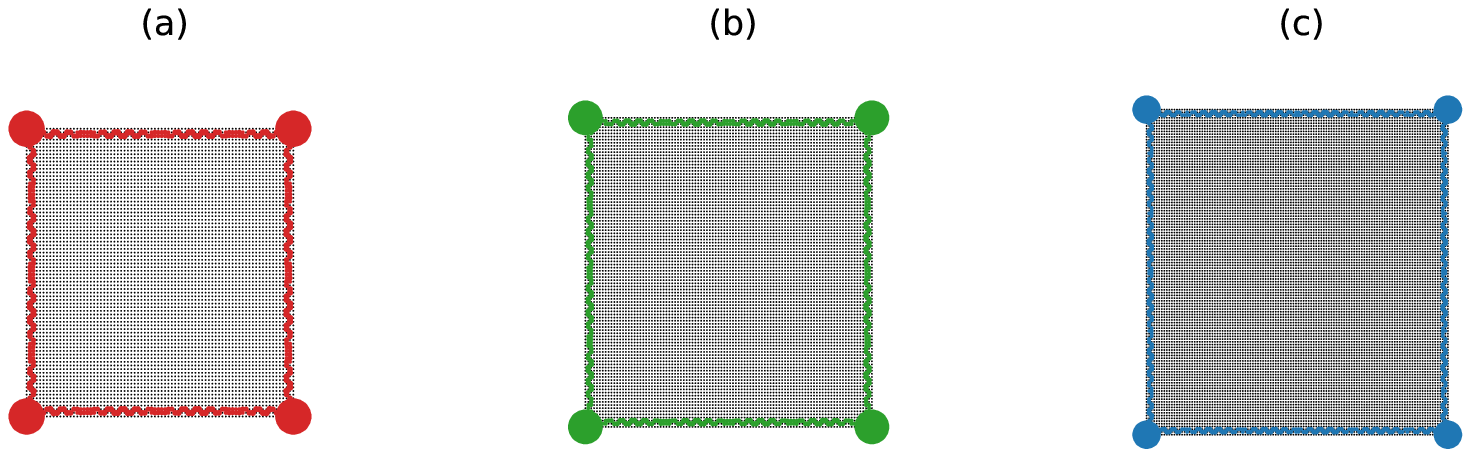}
\caption{(color online). Robustness of the CI state against size effect. We consider various sizes of square lattice with (a) 40$\times$40$\times$4, (b) 50$\times$50$\times$4 and (c) 60$\times$60$\times$4 sites. Here, The hopping and flux parameters are chosen as $v=10.0$, $w^{\prime}=0.75$ and $\phi=-0.25\pi$.}
\label{ACI3}
\end{figure}

In order to check the robustness of the present CI state, we introduce some defects and consider the eigenstate in real-space. Here, we show the energy levels and the distribution of topological state in Fig.~\ref{ACI2}. The isolated corner states, as well the near-edge states, remain robust. Around the defects, the additional corners states and near-edge states emerge, manifesting the robustness of the topological boundary states against the defects [Fig.~\ref{ACI2} (b) and (c)]. The boundary states are also robust against the large enough lattice size, the coexistence of isolate corner states and near-edge states are observed in $40\times 40\times4$, $50\times 50\times4$ and $60\times 60\times 4$ lattices as shown in Fig.~\ref{ACI3}.

\begin{figure}[!htb]
\includegraphics[scale=0.8]{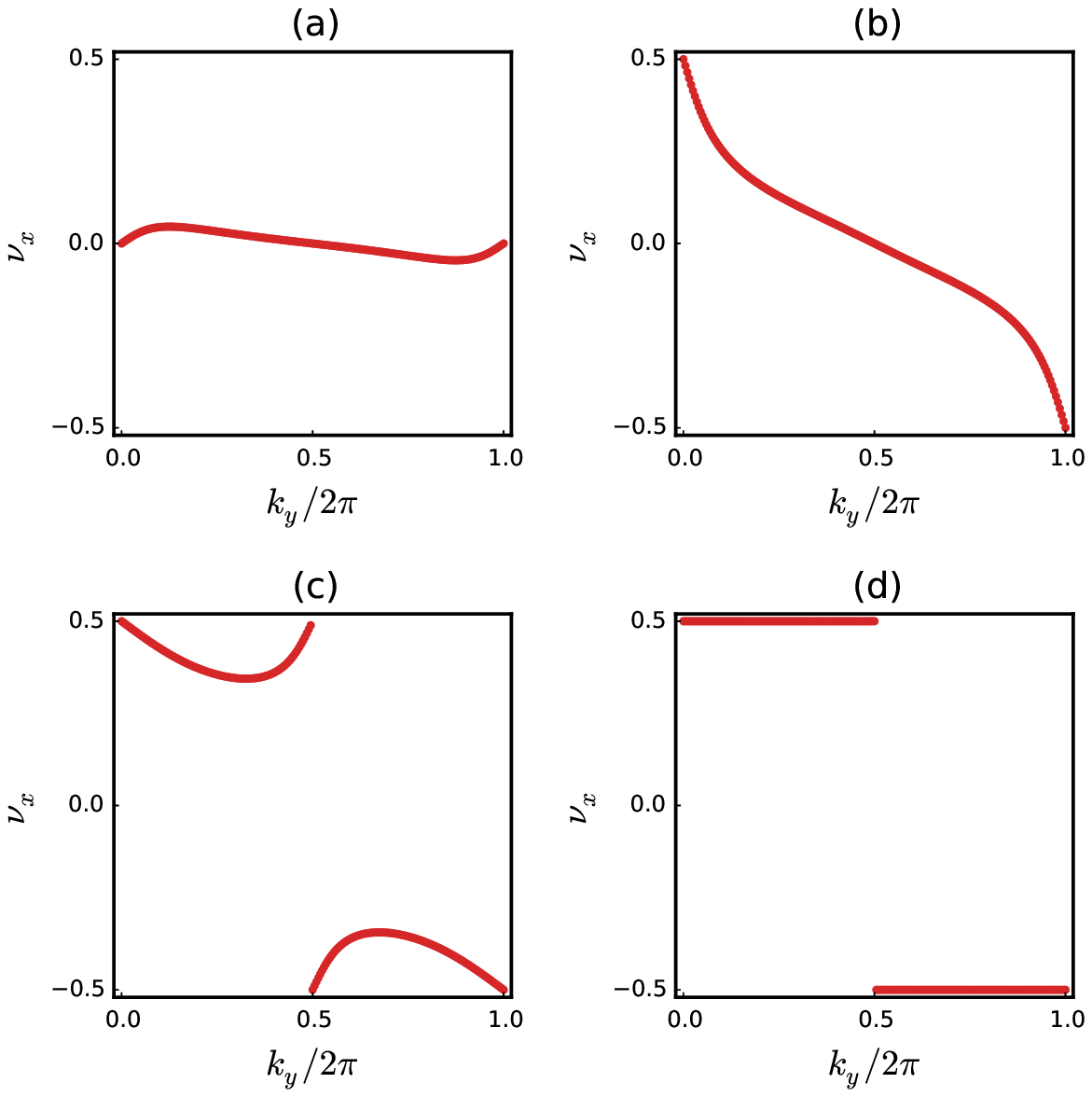}
\caption{(color online). Wannier bands of (a) normal insulators with $v=0.5$, (b) CI-like metallic phase with $v=1.0$ and (c) the second ATI phase with $v=3.0$. In (d), the Wannier bands of $2$D SSH model with $v=3.0$.  The hopping and flux parameters of our model are chosen as $w^{\prime}=0.75$ and $\phi=-0.25\pi$. }
\label{WannC}
\end{figure}

\begin{figure}[!htb]
\includegraphics[scale=0.8]{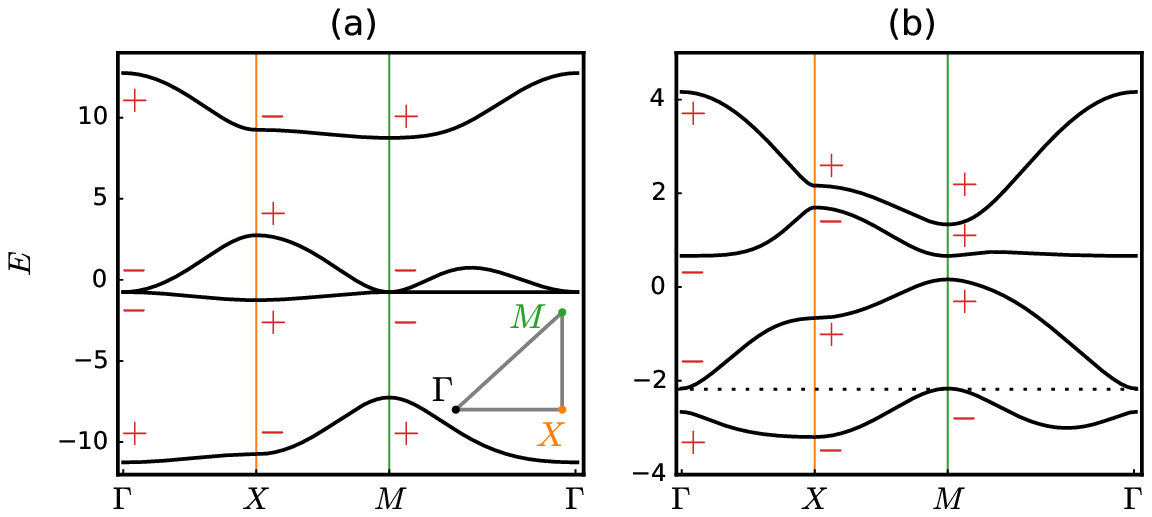}
\caption{(color online). Bulk energy bands along the high symmetric direction at fixed $w^{\prime}=0.75$ (a). $v=5.0$, $\phi=0$, (b). $v=1.0$, $\phi=0.25\pi$. The parity at the high symmetric points ($\Gamma$, $X$ and $M$) is denoted by $\pm$.}
\label{zak_phase}
\end{figure}

\begin{figure}[!htb]
\includegraphics[scale=0.8]{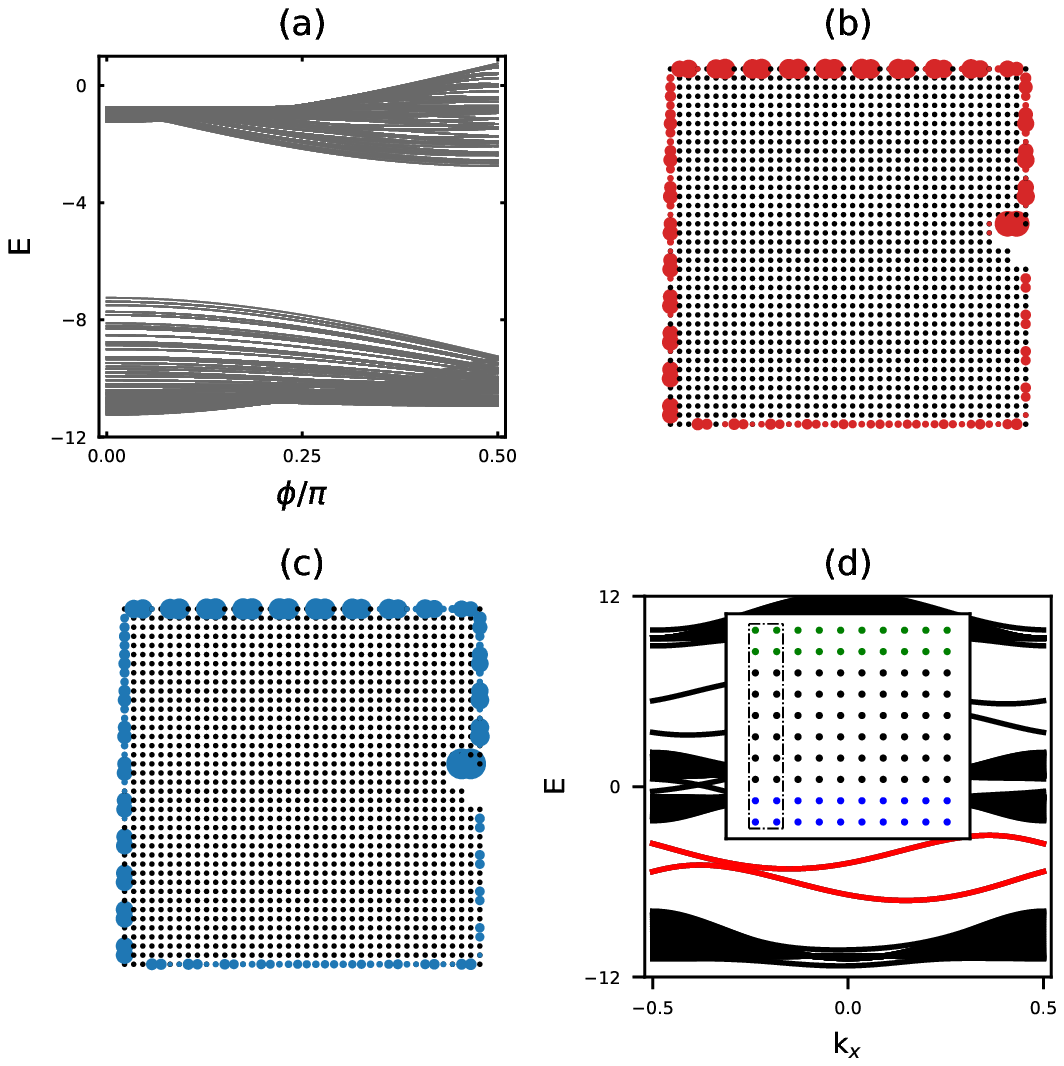}
\caption{(color online). (a). Eigen-energies $E$ versus the staggered flux $\phi$ on the torus geometry. (b) ($v=5.0$) and (c) ($v=10.0$) are the distributions of the zero Berry curvature topological state in a square lattice with defect. (d). The energy band on cylinder geometry with inversion symmetry broken and edge state still emerges at 1/4-filling marked with red. By adding the different on-site potential $V=\pm 1.0$ at sites colored with green and blue to break the inversion symmetry in the inset. The hopping and flux parameters are chosen as $w^{\prime}=0.75$ and $\phi=-0.25\pi$.}
\label{ZBPTI}
\end{figure}

\section{Zero-Chern number topological states} \label{ATI2_WCs}
As the topological invariant, the Wannier bands in the HOTIs are gapped out~\cite{HOTI1,HOTI2,HOTI8}. On the other hand, CIs host a gapless Wannier bands, and the Chern number can be obtained from the Wannier centers~\cite{HOTI2}. We have shown the Wannier centers of conventional CIs and the present CI state in the main text. Here, we show the Wannier bands of the proposed topological state at $1/4$-filling, together with that in the normal insulator state with $v=0.5$, the Chern-like metallic state $v=1.0$. The Wannier centers $\nu(0)=\nu(\pi)=\nu(2\pi)=0$ in the normal insulator and others are near zero [Fig.~\ref{WannC} (a)]. In comparison, the Wannier centers in the metallic CI-like phase are similar to the CIs, where the gapless Wannier bands with $\nu_x(0)=0.5$, $\nu_x(\pi)=0.0$ and $\nu_x(2\pi)=-0.5$ is observed [Fig.~\ref{WannC} (b)]. However, due to the indirect gap between the first and the second bands shown in main text, it is more likely a metallic state. For the zero-Chern number topological state ($v=3.0$), the Wannier bands are gapped out, in agreement with the missing corner states shown in main text. The Wannier centers [Fig.~\ref{WannC} (c)] is similar to a novel topological state reported in the previous $2$D SSH model~\cite{Liu1} [Fig.~\ref{WannC} (d)] with $\nu_x(0)=0.5$, $\nu_x(\pi)=\pm0.5$ and $\nu_x(2\pi)=-0.5$. Hence, we can use these special Wannier bands and Wannier centers to identify various topological phases.

In additional, this zero-Chern number topological state can be identified by the $2$D Zak phase~\cite{Liu1}, which can be implemented based on the parity.
We show the parity at each high symmetric points, marked with $\pm$ in Fig.~\ref{zak_phase}. When $v=5.0$, $w^{\prime}=0.75$ and $\phi=0$, the parity of each band is the same as the topological state in $2$D SSH model~\cite{Liu1}. The parity of the lowest band remains unchanged at $\phi=0.25\pi$. Therefore, no phase transition occurs, which can be further verified with the aid of the evolution of the energy gaps with $\phi$  [Fig.~\ref{ZBPTI}(a)].  Based on Eq.~\ref{Chern_syp}, we obtain the Chern number of the lowest band is zero. However, based on the Eq.~\ref{Zak}, the wave polarization is $(p_x,p_y)=(1/2,1/2)$, which suggest that the $2$D Zak phase is $(\pi,\pi)$. Without lattice dimerization, i.e., $v=1.0$, the parity in $M$-point changes the sign, and the Chern number at $1/4-$filling is ${\cal C}=1$, however, there is no energy gap between the two lowest bands.

We check the eigen-energies of the first and the second energy bands versus the staggered flux $\phi$ on the torus. No band crossing by tuning $\phi$ is observed [Fig.~\ref{ZBPTI}(a)], which indicates no phase transition. Therefore, the topological state at $\phi=0$ observed previously in the $2$D SSH model and the topological state reported here belong to the same topological class, though the time reversal symmetry is preserved, and is broken in the former, and in the latter, respectively. We further show the robustness of this topological state with $v=5.0$ and $v=10.0$ in Fig.~\ref{ZBPTI} (b) and (c). This topological state is very robust for the boundary states bypassing the defect. In addition, we find this topological state is not protected by the inversion symmetry. When we add different on-site potential along the boundary [detail shown in Fig.~\ref{ZBPTI} (d)] to break the inversion symmetry, we find the edge state still emerge at 1/4-filling.

\begin{figure}[!htb]
\includegraphics[scale=0.8]{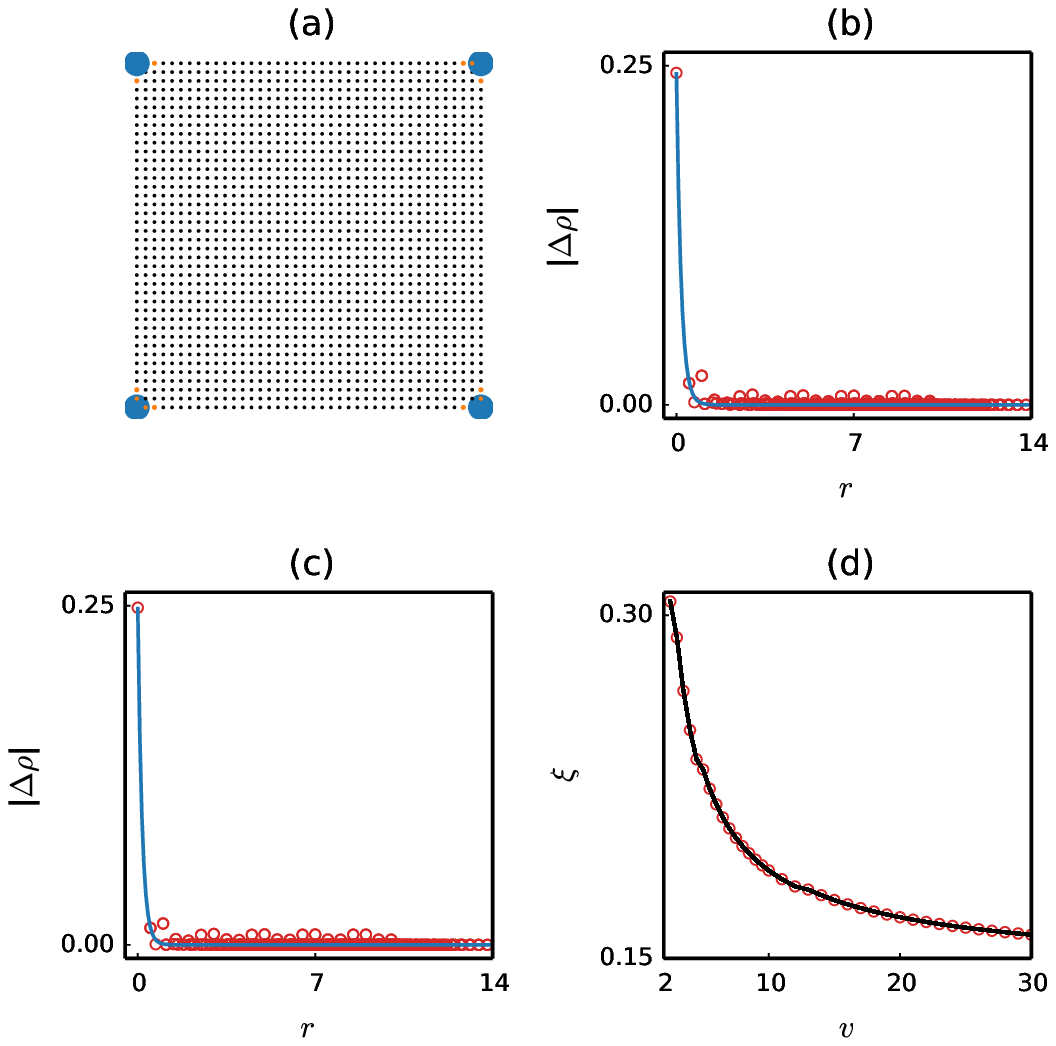}
\caption{(color online). (a). The many-particle density difference at 1/4-filling. Distribution of many particle density difference $\Delta \rho$ in real space with $v=10$ (b) and $v=20$ (c). Here, the fitting function is plotted with blue line. (d) The localized length $\xi$ of fraction charge as a function of intercellular hopping potential $v$.
}
\label{CNFC}
\end{figure}

\section{Robust fractional corner charge} \label{FC_corner}

The many particle density $\rho (r)$ in free fermion systems can be defined as,
\begin{equation}\label{rho_r}
\rho(r) = \sum_{E_n<E_F} |\phi_n({\bf r})|^2,
\end{equation}
where $\phi_n({\bf r})$ is the single-particle state with energy $E_n$ with position ${\bf r}=(x,y)$ of site and $r=|\bf r|$. We also define the many particle density difference $\Delta \rho (r)= \rho(r)-\rho_0(r)$. $\rho_0(r)$ is a standard density, at $\frac{1}{4}$-filling, $\rho_0(r)=1/4$. We numerically find a ?1/4 fractional charge exist around each corner [Fig.~\ref{CNFC} (a)] at quarter filling with open boundary condition. We have also display the many particle density difference $|\Delta \rho (r)|$ in a $20\times20\times4$-sites lattice [in Fig.~\ref{CNFC} (b) and (c)]. Clearly, the many particle density difference near $0.25$ is almost localized around corners, which suggests the appearance of fractional charge. We can use a function $f(r)=A{\rm exp}(-\xi/r)$ to fit $\Delta \rho(r)$ [blue line in Fig.~\ref{CNFC} (b) and (c)]. Here $\xi$ is the localized length which indicates the width of localization. We have also displayed the localized length $\xi$ as a function of intercellular hopping potential $v$ in Fig.~~\ref{CNFC} (d), and the $\xi$ is range from 0.3 to 0.16 (the lattice constant is 1.0), which suggests fractional charge well localized around each corner.

\end{document}